\documentclass[12pt,a4paper]{article}
\usepackage[utf8]{inputenc}  %schaltet Utf8-Kodierung an
\usepackage{cancel}
\usepackage{ucs}             %unicode-Unterstützung
\usepackage[T1]{fontenc}     %Umlaute stehen im Ausdruck als ein Buchstabe
\usepackage{lmodern}         %wird für die Umlaute benötigt
\usepackage{amsmath}         %writing math formulas
\usepackage{amssymb}%superset of the amsfont package, set of msam and msbm fonts
\usepackage{amsthm}
\usepackage{bbm}             %fuer bold math
\usepackage{bm}              %fuer bold math
\usepackage[mathscr]{eucal}  %use of Euler script, Gothic letters with \mathscr
\usepackage{leftidx}         %for left indices
\usepackage{wrapfig}
\usepackage[dvips]{graphicx}
\usepackage[usenames]{color} %Here the option [usenames] causes the definitions
\usepackage{comment}         %Kommentare werden ein- und ausschaltbar
\usepackage{hyperref}       %hyperref zum Schluss
\definecolor{Black}{rgb}{0.,0.,0.}
\definecolor{Blue}{rgb}{0.,0.,1.}
\definecolor{Green}{rgb}{0.,1.,0.}
\definecolor{Cyan}{rgb}{0.,1.,1.}
\definecolor{Red}{rgb}{1.,0.,0.}
\definecolor{Magenta}{rgb}{1.,0.,1.}
\definecolor{Yellow}{rgb}{1.,1.,0.}
\definecolor{White}{rgb}{1.,1.,1.}
\definecolor{Blue4}{rgb}{0.,0.,0.5625}
\definecolor{Blue3}{rgb}{0.,0.,0.6875}
\definecolor{Blue2}{rgb}{0.,0.,0.8125}
\definecolor{LtBlue}{rgb}{0.52734375,0.8046875,1.}
\definecolor{Green4}{rgb}{0.,0.5625,0.}
\definecolor{Green3}{rgb}{0.,0.6875,0.}
\definecolor{Green2}{rgb}{0.,0.8125,0.}
\definecolor{Cyan4}{rgb}{0.,0.5625,0.5625}
\definecolor{Cyan3}{rgb}{0.,0.6875,0.6875}
\definecolor{Cyan2}{rgb}{0.,0.8125,0.8125}
\definecolor{Red4}{rgb}{0.5625,0.,0.}
\definecolor{Red3}{rgb}{0.6875,0.,0.}
\definecolor{Red2}{rgb}{0.8125,0.,0.}
\definecolor{Magenta4}{rgb}{0.5625,0.,0.5625}
\definecolor{Magenta3}{rgb}{0.6875,0.,0.6875}
\definecolor{Magenta2}{rgb}{0.8125,0.,0.8125}
\definecolor{Brown4}{rgb}{0.5,0.1875,0.}
\definecolor{Brown3}{rgb}{0.625,0.25,0.}
\definecolor{Brown2}{rgb}{0.75,0.375,0.}
\definecolor{Pink4}{rgb}{1.,0.5,0.5}
\definecolor{Pink3}{rgb}{1.,0.625,0.625}
\definecolor{Pink2}{rgb}{1.,0.75,0.75}
\definecolor{Pink}{rgb}{1.,0.875,0.875}
\definecolor{Gold}{rgb}{1.,0.83984375,0.}
\newcommand{\ist}[1]{\overset{\footnotesize(\ref{#1})}{=}}

\newcommand{\iist}[2]{\overset{^{(\ref{#1})}}{\underset{^{(\ref{#2})}}{=}}}
\newcommand{\approaches}[1]{\stackrel{\footnotesize #1}{\longrightarrow}}

\theoremstyle{definition}

\theoremstyle{plain}

\begin{document}

\title{A geometric model in 3+1D space-time for electrodynamic phenomena}
\author{Manfried Faber,\\ Technische Universität Wien, Atominstitut\\Operngasse 9, 1040 Wien, Austria}

\maketitle

%------------------------------------------------------------------------------
\begin{abstract}
%------------------------------------------------------------------------------
With the idea to find geometric formulations of particle physics we investigate the predictions of a three dimensional generalisation of the Sine-Gordon model, very close to the Skyrme model and to the Wu-Yang description of Dirac monopoles. With three rotational degrees of freedom of spatial Dreibeins we formulate a Lagrangian and confront the predictions to electromagnetic phenomena. Stable solitonic excitations we compare with the lightest fundamental electric charges, electrons and positrons. Two Goldstone bosons we relate to the properties of photons. These particles are characterised by three topological quantum numbers, which we compare to charge, spin and photon number. Finally we conjecture some ideas for further comparisons with experiments.
\end{abstract}

%------------------------------------------------------------------------------
\section{Introduction}\label{Sec-Intro}
%------------------------------------------------------------------------------
The standard model of particle physics (SM) and our basic theories, quantum field theory (QFT) and quantum mechanics (QM) are extremely successful. A comparison, excellently suited to demonstrate the high accuracy of these theories, leads to the astonishing agreement between theory and experiment for the gyromagnetic ratio of the electron with a precision of better than 12 digits~\cite{Gabrielse:2006}.

On the other hand there are intensive discussions about the interpretation of QM~\cite{BohmDavid2020DBco,SmolinLee2006Ttwp,WoitPeter2006New} and about the mechanisms in nature explaining the phenomena excellently described by the SM. I do not need to repeat here the well-known quote of Feynman about QM. If one reads carefully between the lines of articles and talks of the main builders of the SM, see e.g. the quotes of Veltman and 't Hooft below, one realises that they know very well about the interesting open questions. The SM has about 20 real and 10 integer parameters. One cannot expect that these numbers can be explained within the SM.

Galileo has shown us, how to learn from nature, by experiments analysed with efficient mathematical tools and not by philosophical considerations. There are experiments which may give us some hints in which directions we could try to proceed. Within QM we can describe the properties of spin perfectly, its quantisation and its contribution to angular momentum. It is astonishing, how well we can describe transitions in rotational-vibrational spectra of molecules. But QM does not tell us what is really rotating. The belt-trick and the sphere-trick, see Fig.~\ref{verwirrt}, fascinate us with their strange rotation properties which can be mathematically well-understood within the group of rotations SO(3) and may be related to the properties of spin, see Sect.~\ref{Sec-Spin}.

The Sine-Gordon model and its mechanical pendulum models are inspiring since they have solitons, characterised by the topological quantum number $\Pi_1(S^1)$. These solitons behave like particles. Their mass is realised by three contributions to the energy densities of the field: potential, torsion and kinetic energy. There is a signal velocity $c$ corresponding to the velocity of light. For the mechanical system it is in the order of m/s. Since the Lagrangian is Lorentz invariant with respect to $c$ we can directly observe Lorentz contraction, in a certain sense making the laws of special relativity visible. The sum of the three energy contributions increases with  the gamma-factor corresponding to $c$, as expected for the mass of a relativistic particle. There are two types of solitons interacting like charges. Solitons and antisolitons have exactly the same properties besides their chirality, resembling of the charges of electrons and positrons. Experimentally no differences were found between electrons and positrons, besides the opposite charge. Recent sophisticated experiments show this impressive agreement also for protons and antiprotons~\cite{ Borchert2022} with ``A 16-parts-per-trillion measurement of the antiproton-to-proton charge–mass ratio''. The Sine-Gordon model~\cite{remoissenet:2003} gives the inspiration that particles may be characterised by topological quantum numbers, leading to the quantisation of charge and to such mirror properties of particles and antiparticles. Even experts in particle physics~\cite{Pietschmann2010,Cabaret:2021wmv,Cabaret:2021sdi} raise philosophical questions about the basic entities of QM and QFT. One may see this as an indication that particle physics does not tell us comprehensively enough what particles are. The Sine-Gordon model may give a hint, understandable also for physics laymen.

Interference of electrons in double-slit experiments and of neutrons in interferometers is excellently described by QM. But QM does not give us any understandable idea how objects, registered in electronic detectors as point-like objects, can interfere with each other. To my knowledge there is only one experiment which may give us a first idea how to solve this puzzle, the experiment on bouncing oil-drops invented by Yves Couder~\cite{Couder2005}.

To get further deep understanding of the phenomena in nature, we can try to construct mathematical models reflecting some of the properties of the above mentioned experiments. A simple mathematical model of this type I try to discuss in this article. Its main predictions are condensed in Sect.~\ref{Sec-EdynMaxwell}. Some of the items mentioned there agree with experiment and some to them may disagree. Such a disagreement should motivate modifications improving the applicability of the model.

A further aim of this article is to vote for a possible alternative paradigm of particle physics. It is going back to Einstein, who originally tried to unify gravitation and electromagnetism. This philosophy is summarised in the Aftermath.

In his article ``More is Different'' Philip Anderson~\cite{Anderson:1972} remarked that physical sciences can be ordered according to the degree of their fundamentality. The fundamental theory of matter, the Standard Model of Particle Physics, is a complex theory of many fields and their degrees of freedom. One of its cornerstones is quantum chromo dynamics with $4\cdot8$ gluon field degrees of freedom (dofs) and 4 complex fields for every quark type. History taught us that we can describe electromagnetic phenomena with 4 components of the photon field and 4 complex fields for every fundamentally charged particle with coupled linear equations, with Maxwell and Dirac equations. From mathematics we know that some non-linear equations can be linearised introducing more dofs. This may lead to the conjecture that Maxwell-Dirac theory is a clever linearisation of a non-linear theory with a smaller number of fields. The disadvantage of linearisations may be, that they hide the real dynamics.

A problem which bothered electrodynamics for a long time was the infinity of the self-energy of the electron, as was nicely described by Martinus Veltman in a talk at the 65th Lindau Nobel Laureate Meeting in Konstanz, recorded in https://www.mediatheque.lindau-nobel.org/videos/34703/mar\-tinus-veltman-discovery-higgs-particle. The problem was finally solved by Hendrik Anthony Kramer's idea of renormalisation, by the absorption of infinities in free parameters of the theory. In his talk Veltman commented the subtraction of two infinities, the bare mass of the electron and the electric self-energy of the electron by the believe of the community: ``May be at some future time we know more and we know how to deal with these infinities. May be we find a better theory, where you go to small distances, may be something happens there, but we postpone that problem. All we are going to say is whatever we do, the result for the mass of the electron is what we observe and how that comes about, who cares.'' A similar statement of Gerard 't Hooft one can read in Ref.~\cite{thooft:2021mf}: ``It was something of a shock to realise that renormalisation can end up as an infinite correction to the theory -- and yet, it works." As these statements indicate it may be worthwhile to think about models which give finite results already at the classical level.

One of the big problems of present status of particle physics is the unification of the interactions. The overwhelming majority of the community thinks that gravitation should be quantised. This article tries to argue for the opposite direction, for a geometrical formulation of electrodynamics moving electrodynamics closer to general relativity. This might lead to a geometical motivation of charge quantisation, similiar to the quantisation of Dirac monopoles and revealed a geometrical origin of spin and its quantisation. A model of this type was formulated in 1999~\cite{Faber:1999ia}, but besides an almost one to one copy~\cite{Kouneiher:2016kne} that article was almost completely ignored by the community. Since the author understood since that time several new aspects of the model and since the original presentation was possibly not so easily readable it is time to include the old and new aspects in a revised presentation of the model and compare it to inspiring experiments which have been performed in the last two decades \cite{Couder2005,Bush2015}.

In Sect~\ref{Sec-Formulation} we give the basic definitions of the model, the degrees of freedom (dofs) of the model and the Lagrangian. In Sect~\ref{MonSol} we derive the four types of soliton solutions and calculate their energies. These solitons can be characterised by two topological quantum numbers corresponding to charge and spin quantum numbers. In Sect.~\ref{Sec-EOM} we derive the general equations of motion and the energy momentum tensor. Sect.~\ref{Sec-Edyn} separates charges and their fields, compares their relations to Maxwell equations, derives Coulomb and Lorentz forces, demonstrates the geometrical origin of the U(1) gauge invariance and interprets photons as Goldstone bosons, characterised by a further topological quantum number.  Critical questions are asked in Sect~\ref{Sec-Ideen} and some conjectures added. The conclusion starts with short comparisons to related models and enumerates several agreements and differences to Maxwell's electrodynamics (MEdyn). In the Aftermath the author tries to summarise his understanding in short statements controversial to the present paradigms.

%------------------------------------------------------------------------------
\section{Formulation of the model}\label{Sec-Formulation}
%------------------------------------------------------------------------------
We work in Minkowski space-time and derive the dynamics from a Lagrangian. The only degrees of freedom of the model are local spatial Dreibeins at every event which we define by elements of the group of rotations in 3 dimensions, by an SO(3)-valued scalar field. In Einsteins gravitational theory gravitation is encrypted in the metric tensor, describing a modification of distances and angles in space-time. We aim therefore to describe all long-range properties of nature, i.e. gravitational and electromagnetic, by the properties of space-time. Particles and their fields are formulated by the same degrees of freedom. With the three degrees of freedom of SO(3) we describe electrons and photons and their interaction. Particles are identified by topological quantum numbers of the field of Dreibeins.

Before we specify the dynamics we have to discuss the geometry.

%------------------------------------------------------------------------------
\subsection{Geometry}\label{Sec-Geom}
%------------------------------------------------------------------------------
\subsubsection{Scalar Field}
%------------------------------------------------------------------------------
To simplify the calculations we describe the mentioned rotations by SU(2) elements~\footnote{The three Pauli matrices $-\mathrm i\sigma_k$ obey the same multiplication rules as the three imaginary quaternionic units and the relation
\begin{equation}\label{AlgMult}
(\vec\sigma\vec a)(\vec\sigma\vec b)
=\vec a\vec b+\mathrm i\vec\sigma(\vec a\times\vec b).
\end{equation}
 We are using the vector symbol for the three coefficients of the imaginary quaternionic units and of the su(2) algebra, $\vec q\vec\sigma$ is a short notation for $q_i\sigma_i$. For simplicity we omit the dot for such three dimensional scalar products, if it does not lead to ambiguities. The $\times$ symbol acts always in algebra space.}
\begin{equation}\label{unitquaternions}
Q(x):=q_0(x)-\mathrm i\vec q(x)\vec\sigma\quad\textrm{with}\quad q_0^2+\vec q^{\,2}=1,
\end{equation}
functions of the point $x^\mu$ in 3D+1 space-time. Since SU(2) which is isomorphic to $S^3$ is the double covering group of SO(3), field configurations $\{Q(x)|x\in M_4\}$ differing by a global center transformation, $Q\to-Q$, are identical SO(3) configurations. We would like to emphasise that using in Eq.~(\ref{unitquaternions}) the real unit $1$ and the three imaginary units $-\mathrm i\sigma_k$ we apply the geometry of $R^4$, of the four dimensional euclidean space, to the determination of distances, areas and volumes on the 3-dimensional sphere $q_0^2+\vec q^{\,2}=1$ in 4-dimensions, when we compare them to distances, areas and volumes in $M_4$. Another useful parametrisation of $Q$ is
\begin{equation}\label{nalpha}
q_0=:\cos\alpha,\quad\vec q:=\vec n\sin\alpha,\quad\alpha\in[0,\frac{\pi}{2}],
\quad\vec n^2:=1\quad\Leftrightarrow\quad\vec n\in\mathbb S^2.
\end{equation}
The rotational angle $\omega$ corresponding to the SU(2) matrices $Q$ is related to $\alpha$ by $\vec\omega=2\vec\alpha=2\alpha\vec n$.

%------------------------------------------------------------------------------
\subsubsection{Vector Field=Connection field}
%------------------------------------------------------------------------------
The derivatives
\begin{equation}\label{derivativeQ}
\partial_\mu Q(x)=:-\mathrm i\vec\Gamma_\mu(x)\,\vec\sigma Q(x)
\end{equation}
map the pair $x,\mu$ to tangential vectors on the SU(2)-manifold. The connection coefficients $\vec\Gamma_\mu$ between neighbouring Dreibeins can be expressed by the parameters in Eq.~(\ref{nalpha})
\footnote{\begin{equation}\label{AblGamma}
\partial_\mu QQ^\dagger\ist{unitquaternions}(\partial_\mu q_0-\mathrm i\vec\sigma\partial_\mu\vec q)(q_0+\mathrm i\vec\sigma\vec q)\ist{AlgMult}\underbrace{q_0\partial_\mu q_0+\vec q\,\partial_\mu\vec q}_{0}-\mathrm i\vec\sigma(q_0\partial_\mu\vec q-\vec q\,\partial_\mu q_0+\vec q\times\partial_\mu\vec q)
\end{equation}}
\begin{equation}\label{ConCof}
\vec{\Gamma}_{\mu}\iist{derivativeQ}{AblGamma}q_0\partial_\mu\vec q
-\vec q\,\partial_\mu q_0+\vec q\times\partial_\mu\vec q\ist{nalpha}
\vec n\,\partial_\mu\alpha+\sin\alpha\cos\alpha\,\partial_\mu\vec n
 +\sin^2\alpha\,\vec n\times\partial_\mu\vec n.
\end{equation}
For the description of electro-magnetic phenomena we relate the geometrical quantity $\vec\Gamma_\mu$ to a dual non-abelian vector potential
\begin{equation}\label{dualPot}
\vec C_\mu:=-\frac{e_0}{4\pi\varepsilon_0c_0}\vec\Gamma_\mu,
\end{equation}
where $e_0$ is the fundamental electric charge unit, $\varepsilon_0$ the dielectric constant and $c_0$ the velocity of light.

The derivatives of vectors in algebra space define the covariant derivate
\begin{equation}\begin{aligned}\label{AblBasis}
\partial_\mu\mathbf v(x)&:=\partial_\mu[\vec\sigma Q(x)\vec v(x)]\ist{ConCof}
\underbrace{\vec\sigma\,\partial_\mu Q(x)\vec v(x)}_
{-\mathrm i[\vec v(x)\vec\sigma][\vec\Gamma_\mu(x)\,\vec\sigma]\,Q(x)}
+\vec\sigma Q(x)\,\partial_\mu\vec v(x)
=\\
&\ist{AlgMult}-\mathrm i\,Q(x)\,\vec\Gamma_\mu(x)\vec v(x)
+\vec\sigma[\vec v(x)\times\vec\Gamma_\mu(x)]\,Q(x)+\vec\sigma Q(x)\,\partial_\mu\vec v(x)=\\
&=\underbrace{-\mathrm i\,Q(x)\,\vec\Gamma_\mu(x)\,\vec v(x)}_\textrm{radial}
+\vec\sigma Q(x)[\partial_\mu-\vec\Gamma_\mu(x)\times]\vec v(x)
\end{aligned}\end{equation}

We assume double differentiable scalar fields and apply Schwarz's theorem
\begin{equation}\label{BezSchwarz}
\partial_\mu\partial_\nu Q(x)=\partial_\nu\partial_\mu Q(x).
\end{equation}
From the corresponding expressions
\begin{equation}\begin{aligned}\label{diffAbls}
\partial_\mu\partial_\nu Q(s,t)&\ist{derivativeQ}
\partial_\mu(-\mathrm i\vec\sigma\vec\Gamma_\nu Q)
=(-\mathrm i\vec\sigma\vec\Gamma_\nu)(-\mathrm i\vec\sigma\vec\Gamma_\mu)Q
-\partial_\mu\vec\Gamma_\nu\mathrm i\vec\sigma Q\\
&\ist{AlgMult}-\vec\Gamma_\nu\vec\Gamma_\mu Q
-(\partial_\mu\vec\Gamma_\nu+\vec\Gamma_\nu\times\vec\Gamma_\mu)\mathrm i\vec\sigma Q,\\
\partial_\nu\partial_\mu Q(s,t)&=-\vec\Gamma_\mu\vec\Gamma_\nu Q
-(\partial_\nu\vec\Gamma_\mu+\vec\Gamma_\mu\times\vec\Gamma_\nu)\mathrm i\vec\sigma Q,
\end{aligned}\end{equation}
we get the Maurer-Cartan equation
\begin{equation}\label{MaurerCartan}
\partial_\mu\vec\Gamma_\nu-\partial_\nu\vec\Gamma_\mu\iist{BezSchwarz}{diffAbls}
2\vec\Gamma_\mu\times\vec\Gamma_\nu,
\end{equation}
which describes that the $4\cdot3$ vector field components are derived from a unique SU(2) scalar field with 3 degrees of freedom only.

%------------------------------------------------------------------------------
\subsubsection{Curvature field}
%------------------------------------------------------------------------------
Unit squares in $M_4$ are mapped to parallelograms in the tangential space of $S^3$ with the area vector
\begin{equation}\label{RSU2}
\vec R_{\mu\nu}:=\vec\Gamma_\mu\times\vec\Gamma_\nu\ist{MaurerCartan}
\frac{1}{2}(\partial_\mu\vec\Gamma_\nu-\partial_\nu\vec\Gamma_\mu).
\end{equation}
Using these two expressions for $\vec R_{\mu\nu}$ we can write it in a more general form
\begin{equation}\label{RAllg}
\vec R_{\mu\nu}\ist{RSU2}\partial_\mu\vec\Gamma_\nu-\partial_\nu\vec\Gamma_\mu
-\vec\Gamma_\mu\times\vec\Gamma_\nu.
\end{equation}
which reminds of the components of the (gauge covariant) field strength tensor in SU(2)-QCD and of the curvature tensor in general relativity. \footnote{At a first glance readers may draw the wrong conclusion that $\partial_\mu QQ^\dagger$ in Eq.~(\ref{derivativeQ}) defines a trivial connection. A trivial connection $\vec A_\mu$ one would get by $\partial_\mu QQ^\dagger=:-\mathrm i\vec A_\mu(x)\,\frac{\vec\sigma}{2}$, differing slightly from Eq.~(\ref{derivativeQ}). Inserting the relation between $\vec A_\mu$ and $\vec\Gamma_\mu$, $\vec A_\mu=2\vec\Gamma_\mu$, to Eq.~(\ref{MaurerCartan}) shows that the two-form
\begin{equation}\label{vanFieldStrength}
\partial_\mu\vec A_\nu-\partial_\nu\vec A_\mu-\vec A_\mu\times\vec A_\nu
\ist{MaurerCartan}0
\end{equation}
is vanishing identically and therefore useless for a determination of areas on $S^3$.}

Here we can also observe the geometrical realisation of gauge transformations, $\vec\Gamma_\mu\to\vec\Gamma_\mu^\prime$, as basis changes in the three dimensional euclidean tangential space at $Q$. $\vec R_{\mu\nu}\vec\sigma Q$ is a vector in this three dimensional tangential space at $Q$ on $S^3$, represented by the special basis vectors $\sigma_i Q$. In this special gauge, the algebraic expression for $\vec R_{\mu\nu}$ is defined by Eq.~(\ref{RSU2}). After a basis change the affine connection is modified, $\vec\Gamma_\mu\to\vec\Gamma_\mu^\prime$, and Eq.~(\ref{RSU2}) cannot be used any more. As one can see, performing the basis change explicitly, the more general Eq.~(\ref{RAllg}) is still valid after arbitrary basis rotations in the tangential space at $Q$, after arbitrary local gauge transformations.

We relate the curvature field $\vec R_{\mu\nu}$ to dual non-abelian field strength tensor components by
\begin{equation}\begin{aligned}\label{FDvonR}
-\frac{e_0}{4\pi\varepsilon_0 c_0}\vec R_{\mu\nu}=:
\frac{1}{2}\varepsilon_{\mu\nu\rho\sigma}
\vec F^{\rho\sigma}=:{^\star}\hspace{-0.8mm}\vec F_{\mu\nu}
=\begin{pmatrix}
0&\vec B_1&\vec B_2&\vec B_3\\
-\vec B_1&0 &\frac{\vec E_3}{c_0}&-\frac{\vec E_2}{c_0}\\
-\vec B_2&-\frac{\vec E_3}{c_0}&0&\frac{\vec E_1}{c_0}\\
-\vec B_3&\frac{\vec E_2}{c_0}&-\frac{\vec E_1}{c_0}&0\end{pmatrix}.
\end{aligned}\end{equation}
where we used the convention $\varepsilon^{0123}=1$ for the total antisymmetric epsilon tensor.

It may be astonishing, why we decided to use this dual representation to describe electric charges, though a short calculation shows that the abelian $\vec{j}_\mathrm{el}^\nu$ and the non-abelian electric current $\vec{J}_\mathrm{el}^\nu$ vanish
\begin{align}\label{ElCurrCons}
\vec{j}_\mathrm{el}^\nu&:=\partial_\mu\vec F^{\mu\nu}\propto
\epsilon^{\mu\nu\rho\sigma}\partial_\mu\vec R_{\rho\sigma}\ist{AblBasis}
\underbrace{\epsilon^{\mu\nu\rho\sigma}\partial_\mu\partial
_\rho\vec\Gamma_\sigma}_{0}=0.\\\label{CovElCurr}
\vec{J}_\mathrm{el}^\nu&:=D_\mu\vec F^{\mu\nu}\propto
\epsilon^{\mu\nu\rho\sigma}D_\mu\vec R_{\rho\sigma}\iist{AblBasis}{RSU2}
\underbrace{\epsilon^{\mu\nu\rho\sigma}\partial_\mu\partial_\rho\vec
\Gamma_\sigma}_{\vec{j}_\mathrm{el}^\nu}-\frac{1}{2}\epsilon^{\mu\nu\rho\sigma}
\Gamma_\mu\times\vec R_{\rho\sigma}=\\\nonumber&=\underbrace
{\epsilon^{\mu\nu\rho\sigma}\partial_\mu\partial_\rho}_{0}\vec\Gamma_\sigma
-\frac{1}{2}\epsilon^{\mu\nu\rho\sigma}\underbrace{
\Gamma_\mu\times(\vec\Gamma_\rho\times\vec\Gamma_\sigma)}_{\vec\Gamma_\rho
(\vec\Gamma_\mu\vec\Gamma_\sigma)-\vec\Gamma_\sigma(\vec\Gamma_\mu\vec\Gamma_\rho)}
=0.
\end{align}
The zeroth component of these equations forbids the existence of non-abelian electric sources
\begin{equation}
D_i\vec E_i\ist{CovElCurr}\partial_i\vec E_i\ist{CovElCurr}0.
\end{equation}
But this does not mean that the abelian electric flux lines are closed. They have sources, the topological solitons of this model.

Already here we would like to mention that in this description with dual vector fields static charges are formulated with spatial components of static vector fields~(\ref{ConCof}). For moving charges and for their magnetic fields ${^\star}\hspace{-0.8mm}\vec F_{0i}\propto\vec R_{0i}$ non-vanishing time-components of the connection $\vec\Gamma_\mu$ are necessary.

After the definition of the geometrical quantities we are ready to define the dynamics.

%------------------------------------------------------------------------------
\subsection{Lagrangian}
%------------------------------------------------------------------------------
To keep solitons stable we need besides the almost familiar term $\mathcal L_\textrm{curv}$, quadratic in the field strength, a potential term $\mathcal  L_\textrm{pot}$
\begin{equation}\label{Lagr4D}
\mathcal L:=\mathcal L_\textrm{curv}+\mathcal L_\textrm{pot}
:=-\frac{\alpha_f\hbar c_0}{4\pi}\left(\frac{1}{4}\,\vec R_{\mu\nu}\vec R^{\mu\nu}
+\Lambda(q_0)\right)\quad\textrm{with}\quad\Lambda(q_0):=\frac{q_0^6}{r_0^4}.
\end{equation}
The prefactor of $\mathcal L_\textrm{curv}$ we have formulated with Sommerfeld's fine structure constant $\alpha_f:=\frac{e_0^2}{4\pi\varepsilon_0\hbar c_0}$[SI] in order to use an expression which is valid in SI and in the cgs unit system. In more familiar terms the square term reads in SI units

\begin{equation}\label{Ldual}
\mathcal L_\textrm{curv}
\iist{Lagr4D}{FDvonR}-\frac{1}{4\mu_0}
{^\star}\hspace{-0.8mm}\vec F_{\mu\nu}{^\star}\hspace{-0.8mm}\vec F^{\mu\nu}
\ist{FDvonR}-\frac{\epsilon_0}{2}\vec E\vec E+\frac{1}{2\mu_0}\vec B\vec B,
\end{equation}
with $\varepsilon_0\mu_0c_0^2=1$. As we will discuss in Sect.~\ref{MonSol} this Lagrangian allows for 4 types of stable solitons with long range, Coulombic interaction. Since we want to compare the properties of these solitons with experiments, we compare them to the lightest monopoles existing in nature, with electrons. This is the reason why we relate $\vec R_{\mu\nu}$ to ${^\star}\hspace{-0.8mm}\vec F_{\mu\nu}$. The sign of $\mathcal L_\textrm{curv}$ does not matter for a classical system. It decides whether quantum fluctuations are suppressed.

We would like to mention that up to the multiplicative constant $\alpha_f$, $\mathcal L_\textrm{curv}$ is geometrically identical to the Skyrme term in the Skyrme model~\cite{Skyrme:1958vn,Skyrme:1961vq}, with the distinction that the Skyrme model is using the vanishing field strength tensor~(\ref{vanFieldStrength}) and describes solitons with short range interaction which are interpreted as baryons. 

$\mathcal L_\textrm{curv}$ tends to expand solitons. To stabilise solitons we need a further term in the Lagrangian, a compression term with a smaller number of derivatives. A term, quadratic in the derivatives, as used by Skyrme, forbids solitons with long range interaction. A basic idea of our model is to have solitons with a hedgehog field extending to infinity and covering only half of $S^3$. This can only be achieved with a potential term $\Lambda(q_0)$ without derivatives and the minimum at $q_0=0$, at the equator of $S^3$. $\Lambda(q_0)=q_0^6/r_0^4$ turned out to lead to the simplest solution, and it is appealing to believe that the simplest solution is also the most natural, even if one does not know yet its origin. The factor $r_0^{-4}$ is necessary by dimensional reasons. $r_0$ is the length scale responsible for the size and the mass of solitons. Fixing $q_0$ to 0 keeps the two dynamical degrees of freedom encoded in the $\vec n$-field, see Eq.~(\ref{nalpha}), and leads to a symmetry broken vacuum. It allows for two massless excitations which can be related to the two massless excitations of electrodynamics, to photons, see Sect.~\ref{Sec-Gold}.

This definition of the Lagrangian leads to three topological quantum numbers, $\pi_2(\mathbb S^2)$, $\pi_3(\mathbb S^3)$ and $\pi_3(\mathbb S^2)$ which we will relate to the topological quantisation of electric charge, spin and photon number, as will be discussed in Sects.~\ref{Sec-Qnum} and \ref{Sec-Gold}.

We are looking for stable static solutions of the equations of motion. The two terms in the Lagrangian~(\ref{Lagr4D}) differ in the number of derivatives leading to a difference in the scale dependence, $\vec x\to\lambda\vec x$
\begin{equation}\begin{aligned}\label{HBeitr}
&H_\mathrm{cur}:\ist{Ldual}\frac{\epsilon_0}{2}\int\mathrm d^3x\vec E_i\vec E_i
\hspace{4mm}\quad\rightarrow\quad
\frac{1}{\lambda}H_\mathrm{cur},\\
&H_\mathrm{pot}:\ist{Lagr4D}\frac{\alpha_f\hbar c_0}{4\pi}\int\mathrm d^3x
\frac{q_0^6}{r_0^4}\quad\rightarrow\quad\lambda^3H_\mathrm{pot}.
\end{aligned}\end{equation}
For stable minima of the total energy
\begin{equation}\label{GesamtE}
H_\mathrm{tot}:=H_\mathrm{cur}+H_\mathrm{pot}
\end{equation}
we get therefore a relation between the corresponding static energy contributions
\begin{equation}\label{EnergieBez}
\frac{\mathrm d}{\mathrm d\lambda}H_\mathrm{tot}\Bigr\rvert_{\lambda=1}
\ist{GesamtE}\frac{\mathrm d}{\mathrm d\lambda}H_\mathrm{cur}\Bigr\rvert
_{\lambda=1}+\frac{\mathrm d}{\mathrm d\lambda}H_\mathrm{pot}\Bigr\rvert_{\lambda=1}
\ist{HBeitr}-H_\mathrm{cur}+3H_\mathrm{pot}=0.
\end{equation}
In Sect.~\ref{MonSol} we can check the validity of this application of the Hobart-Derrick theorem~\cite{Hobart:1963rh,Derrick:1964gh} for the analytical solution.

%------------------------------------------------------------------------------
\section{Soliton solutions}\label{MonSol}
%------------------------------------------------------------------------------
\subsection{Hedgehog solution}\label{Sec-hedgehog}
%------------------------------------------------------------------------------
We expect to find a spherical symmetric soliton with the shape of a hedgehog.\\[2mm]
\begin{minipage}[c]{65mm}
\includegraphics[scale=0.31]{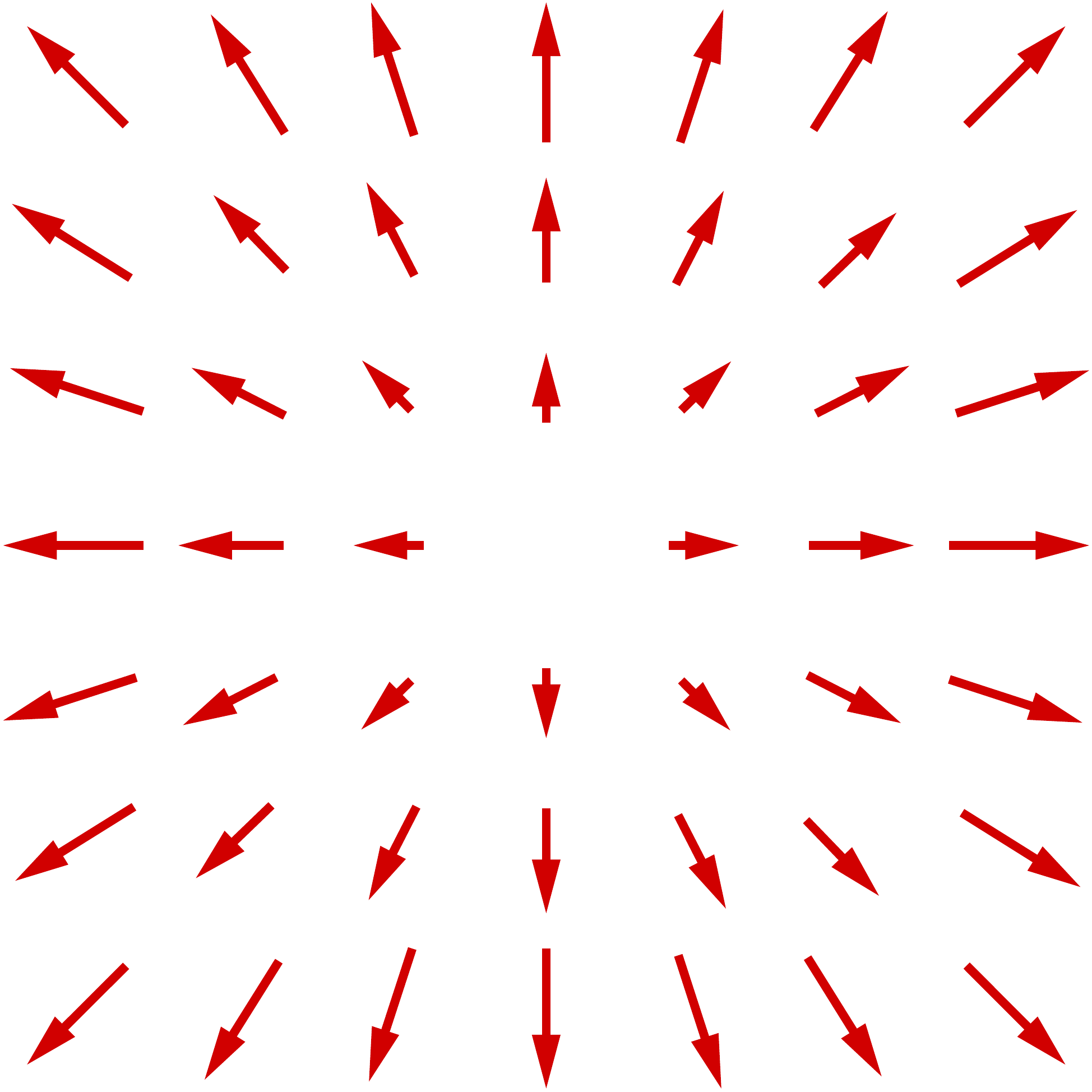}
\label{schemaigel}
\end{minipage}
\begin{minipage}[c]{70mm}
\vspace{2mm} We use the parameters~(\ref{nalpha}) to define the spherically symmetric hedgehog
\begin{equation}\begin{aligned}\label{RegularIgel}
&n_i(x):=\frac{x_i}{r},\\
&\alpha(r)\in[0,\frac{\pi}{2}]\\
&\alpha(0)=0,\quad\alpha(\infty)=\pi/2.
\end{aligned}\end{equation}
Depicted is the imaginary part
$$\vec q(x)=\vec n(x)\sin\alpha(x)$$
of the soliton field $Q(x)$ of Eq.~(\ref{unitquaternions}) in a symmetry plane.\\
\end{minipage}\\
From the soliton field $Q(x)$ we get the connection coefficients
\begin{equation}\begin{split}\label{sphConCof}
\vec\Gamma_r&\ist{ConCof}\vec n\,\partial_r\alpha,\\
\vec\Gamma_\vartheta&\ist{ConCof}
\sin\alpha\left[\cos\alpha\,\vec e_\theta+\sin\alpha\,\vec e_\phi\right]
 =:\sin\alpha\,\vec e_\xi,\\
\vec\Gamma_\varphi&\ist{ConCof}\sin\vartheta\sin\alpha
\left[\cos\alpha\,\vec e_\phi-\sin\alpha\,\vec e_\theta\right]
=:\sin\vartheta\sin\alpha\,\vec e_\eta,
\end{split}\end{equation}
where $\vec n,\vec e_\theta,\vec e_\phi$ and $\vec e_\xi:=\cos\alpha\,\vec e_\theta+\sin\alpha\,\vec e_\phi$, $\vec e_\eta:=\cos\alpha\,\vec e_\phi-\sin\alpha\,\vec e_\theta$ are spherical unit vectors in the su(2) algebra.

For the determination of the dual vector potential from the geometrical quantity $\vec\Gamma_\mu$ we have to take into account the length scales in  spherical coordinates
$(l_r,l_\vartheta,l_\varphi)=(1,r,r\sin\vartheta)$
\begin{equation}\label{sphericalVecPots}
\vec C_r\ist{dualPot}-\frac{e_0}{4\pi\varepsilon_0c_0}\frac{\vec\Gamma_r}{l_r},\quad
\vec C_\vartheta
=-\frac{e_0}{4\pi\varepsilon_0c_0}\frac{\vec\Gamma_\vartheta}{l_\vartheta},\quad
\vec C_\varphi
=-\frac{e_0}{4\pi\varepsilon_0c_0}\frac{\vec\Gamma_\varphi}{l_\varphi}.
\end{equation}

\label{vecfieldcomp}We would like to mention, that due to the dual formulation and in distinction to Maxwell's electrodynamics (MEdyn) only the spatial components $\vec\Gamma_i, i\in \{1,2,3\}$ of the vector field $\vec\Gamma_\mu$ are involved in the description of static charges. In this way static problems are described with static fields and spatial derivatives. Time dependent problems need time dependent fields and time derivatives.

Up to the measure system depending constant $-\frac{e_0}{4\pi\varepsilon_0}$, the electric field strength is given by the ratio of an area in the tangential space of $S^3$ divided by the corresponding area at $(r,\vartheta,\varphi)$ in 3-dimensional (3D) space
\begin{alignat}{3}
&\vec E_r\iist{RSU2}{FDvonR}-\frac{e_0}{4\pi\varepsilon_0}
\frac{\vec\Gamma_\vartheta\times\vec\Gamma_\varphi}{r^2\sin\vartheta}
\ist{sphericalVecPots}-\frac{e_0}{4\pi\varepsilon_0}\frac{\sin^2\alpha}{r^2}
\vec n\quad&&\approaches{r\to\infty}\quad
&&-\frac{e_0}{4\pi\varepsilon_0}\frac{1}{r^2}\,\vec n,\nonumber\\
\label{EFieldStrengthHedgeHog}
&\vec E_\vartheta=-\frac{e_0}{4\pi\varepsilon_0}
\frac{\vec\Gamma_\varphi\times\vec\Gamma_r}{r\sin\vartheta}
=-\frac{e_0}{4\pi\varepsilon_0}\,\frac{\sin\alpha\,\partial_r\alpha}{r}
\,\vec e_\xi\quad&&\approaches{r\to\infty}\quad&&\hspace{10mm}0,\\
&\vec E_\varphi=-\frac{e_0}{4\pi\varepsilon_0}
\frac{\vec\Gamma_r\times\vec\Gamma_\vartheta}{r}
=-\frac{e_0}{4\pi\varepsilon_0}\,\frac{\sin\alpha\,\partial_r\alpha}{r}
\,\vec e_\eta\quad&&\approaches{r\to\infty}\quad&&\hspace{10mm}0.\nonumber
\end{alignat}
From $\alpha(r)\approaches{r\to\infty}\pi/2$ we get $\sin\alpha(r)\to 1$ and $\partial_r\alpha(r)\to 0$. Only the radial electric field strength survives for large $r$ and approaches zero like $r^{-2}$. Its abelian part gets asymptotically the intended result~\footnote{With bold characters we indicate vectors in 3D.}
\begin{equation}
\vec{\mathbf E}\vec n\approaches{r\to\infty}-\frac{1}{4\pi\varepsilon_0}\frac{e_0}{r^2}\,\frac{\mathbf r}{r}.
\end{equation}

The shape of the profile function $\alpha(r)$ follows from a minimisation of the total energy
\begin{align}
H_\mathrm{tot}&\iist{GesamtE}{nalpha}
\int{\mathrm d^3 r}\left[\frac{\varepsilon_0}{2}(\vec E_r\vec E_r
+\vec E_\vartheta\vec E_\vartheta+\vec E_\varphi\vec E_\varphi)+
\frac{\alpha_f\hbar c_0}{4\pi}\frac{\cos^6\alpha}{r_0^4}\right]=\nonumber\\
&\ist{EFieldStrengthHedgeHog}\alpha_f\hbar c_0\int_0^\infty{\mathrm dr}
\left[\frac{\sin^4\alpha}{2 r^2} +(\partial_r\cos\alpha)^2
+\frac{r^2}{r_0^4}\cos^6\alpha\right]=\label{totEneMonop}\\
&=\frac{\alpha_f\hbar c_0}{r_0}\int_0^\infty{\mathrm d\rho}
\left[\frac{\sin^4\alpha}{2\rho^2}+(\partial_\rho\cos\alpha)^2
+\rho^2\cos^6\alpha\right]\quad\textrm{with}\quad\rho:=\frac{r}{r_0},\nonumber
\end{align}
where $r_0$ defines the scale for the size and the mass of the soliton. The first term in the final energy expression is due to the radial electric field, the second due to the two tangential fields and the third due to the potential energy.

\begin{figure}[h!]
\centering
\input{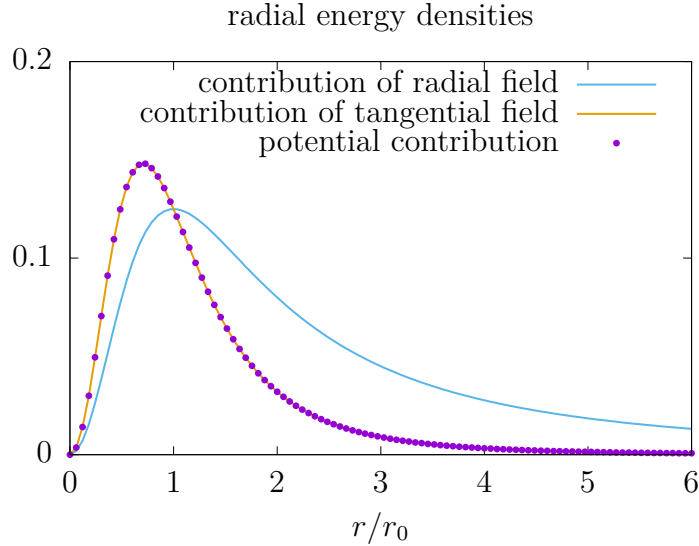}
\caption{Radial energy densities of the stable soliton solution according to Eq.~(\ref{radialdensities}). The prefactor in $h$ is omitted.}
\label{enedens3}
\end{figure}

By the variation of the energy functional~(\ref{totEneMonop}) with respect to $\alpha(\rho)$ we get the differential equation for the minimum of the energy
\begin{equation}\label{nlDE}
\frac{(1-\cos^2\alpha)\cos\alpha}{\rho^2}+\partial^2_\rho\cos\alpha
-3\rho^2\cos^5\alpha=0,
\end{equation}
a nonlinear differential equation of second order. This equation is fulfilled by
\begin{equation}\label{MinLoesm3}
\alpha\ist{nlDE}\arctan\rho=\arctan\frac{r}{r_0} .
\end{equation}
With 
\begin{equation}\label{sincos}
\sin\alpha(r)=\frac\rho{\sqrt{1+\rho^2}}\qquad
\cos\alpha(r)=\frac{1}{\sqrt{1+\rho^2}} ,
\end{equation}
we get the three contributions to the radial energy density, the $r$ dependent spherical integrals in Eq.~(\ref{totEneMonop}),
\begin{equation}\label{radialdensities}
h\ist{totEneMonop}\frac{\alpha_f\hbar c_0}{r_0}\Big[\frac{\rho^2}{2(1+\rho^2)^2}
 +\frac{\rho^2}{(1+\rho^2)^3}+\frac{\rho^2}{(1+\rho^2)^3}\Big]
\end{equation}
listed in the same order as in Eq.~(\ref{totEneMonop}). They are depicted in Fig.~\ref{enedens3} without the prefactor $\frac{\alpha_f\hbar c_0}{r_0}$.
The integrals over these radial energy densities behave as 2:1:1 and lead to the total energy
\begin{equation}\label{EneMonopFormel}
H_\mathrm{tot}=\frac{\alpha_f\hbar c_0}{r_0}\frac{\pi}{4}.
\end{equation}
The ratio 2:1 between radial and tangential energies is specific to the chosen power in the potential energy. The ratio 4:1 between total energy and potential energy contribution is due to the Hobart-Derrick theorem, see Eq.~(\ref{EnergieBez}). We can try to fix the parameter $r_0$ of the model by a comparison with experiments. The only known fundamental stable monopoles are electrons with a self-energy of 0.511~MeV. This comparison leads to
\begin{equation}\label{r0m2}
r_0=2.21~\mathrm{fm}.
\end{equation}
It is in the order of the classical electron radius $r_\mathrm{cl}:=\frac{\alpha_f\hbar c_0}{m_ec_0^2}=$~2.82~fm.

The stable soliton configuration, derived in this section, is tightly related to the monopoles which Dirac invented~\cite{dirac:1931kp,dirac:1948um}, to describe magnetic monopoles within MEdyn. Dirac monopoles have two types of singularities, a Dirac string which is necessary to allow for a non-vanishing magnetic charge and a singularity in the center well-known for all point-like charges in MEdyn. Due to the presence of the quantised string, Dirac monopoles have quantised magnetic charges. Wu and Yang~\cite{wu:1975vq,wu:1976qk} found another representation of point-like Dirac monopoles by a normalised $\vec n$-field of the same structure as defined in Eq.~(\ref{RegularIgel}) where the Dirac string is removed and the singularity in the center is still existing. This point-like singularity warrants the quantisation of magnetic charges. In this article we emphasize that one should regularise this central singularity by enlarging the degrees of freedom to an SU(2) scalar field and define a Lagrangian which stabilises such monopole configurations. With these modifications and a dual interpretation of the fields one gets a classical field theory for Dreibeins in space-time which has many features well-known from classical electrodynamics. It is an interesting subject to investigate which predictions of this dual formulation of electrodynamics differ from Maxwell's electrodynamics (MEdyn) and which predictions disagree with the experiments to electrodynamics (Edyn).

%------------------------------------------------------------------------------
\subsection{Quantum numbers of solitons}\label{Sec-Qnum}
%------------------------------------------------------------------------------
The topological structure of solitons is characterised by discrete, integer or half-integer topological quantum numbers. They lead to integer charges, half-integer spin quantum numbers and ineger photon numbers and to the astonishing mirror properties of particles and antiparticles.

%------------------------------------------------------------------------------
\subsubsection{Charge}\label{Sec-Ladung}
%------------------------------------------------------------------------------
One can classify field configurations mapping spheres to spheres. Spatial infinity is isomorphic to $\mathbb S^2$. The field of a single soliton at the spatial infinity $\mathbb S^2_\infty$ is characterised by $q_0=0$ and varying $\vec n$-field depending on the spatial direction and covering the equatorial $\mathbb S^2_\mathrm{equ}$ of $\mathbb S^3$. The map $\Pi_2(\mathbb S^2):\mathbb S^2_\infty\mapsto\mathbb S^2_\mathrm{equ}$ of Eq.~(\ref{RegularIgel}) is characterised by the winding number $Z=1$ which we can relate to the electric charge $Q_\mathrm{el}$ of the soliton by
\begin{equation}\label{Qund Z}
Q_\mathrm{el}:=-e_0Z.
\end{equation}
In the limit $r_0\to0$ an arbitrary smooth SU(2)-valued scalar field reduces to a scalar $\vec n$-field with $\vec n^2=1$, i.e an $\mathbb S^2$-field, with Dirac monopoles in the Wu-Yang representation~\cite{wu:1975vq,wu:1976qk}. This justifies to define the charge number $Z$ enclosed in a connected volume $V$ with a surface $\partial V$ isomorphic to $\mathbb S^2$ by the topological quantum number associated to the map $\Pi_2(\mathbb S^2)$
\begin{equation}\label{GaussFluss}
Z(V):=\frac{1}{4\pi}\oint_{\mathbb S(u,v)}\mathrm du\mathrm dv\,
\vec n[\partial_u\vec n\times\partial_v\vec n]
\end{equation}
where we parametrised the surface by two parameters $u$ and $v$, e.g. $\vartheta$ and $\varphi$.

%------------------------------------------------------------------------------
\subsubsection{Topological charge}\label{Sec-TopLad}
%------------------------------------------------------------------------------
The hedgehog solution found in Sect.~\ref{Sec-hedgehog} is characterised by a further quantum number, the number of coverings of the SU(2) manifold, the topological charge
\begin{equation}\label{TopolChar}
\mathcal Q:=\frac{1}{2\pi^2}
\int_0^\infty\mathrm dr\int_0^\pi\mathrm d\vartheta\int_0^{2\pi}\mathrm d\varphi\,
\vec\Gamma_r(\vec\Gamma_\vartheta\times\vec\Gamma_\varphi)
=\frac{1}{2\pi^2}\int\mathrm d^3r\;\vec\Gamma_x(\vec\Gamma_y\times\vec\Gamma_z).
\end{equation}
$\vec\Gamma_x(\vec\Gamma_y\times\vec\Gamma_z)$ is here the Jacobi determinant for the map of the three dimensional space to $\mathbb S^3$. $2\pi^2$ is the volume of $\mathbb S^3$, therefore $\mathcal Q=1$ describes a full covering. The hedgehog configuration~(\ref{RegularIgel}) covers only half of $\mathbb S^3$ resulting in $\mathcal Q=\frac{1}{2}$.

By Poincaré transformations we remain in the same topological sector as long as we restrict them to translations, rotations and boosts, since they belong to the connected component of the identity. By parity transformations $\Pi_n$ and transformations with the non-trivial center element $z=-1$ of SU(2)
\begin{align}\label{Pin}
&\Pi_n:\quad \vec n\to-\vec n,\quad Q=q_0-\mathrm i\vec n\vec\sigma\sin\alpha\to Q^\dagger=q_0+\mathrm i\vec n\vec\sigma\sin\alpha\quad\forall x\\\label{ZentrTrans}
&z:\quad Q\to -Q\quad\forall x\
\end{align}
we can get topologically stable monopole configurations differing in the topological quantum numbers $Z$ and $\mathcal Q$, enumerated in Table~\ref{TabSign}.
\begin{table}[h]\begin{center}\hspace*{-10mm}\begin{tabular}{cccc}\hline 
$\mathcal T=1$&$\mathcal T=z$&$\mathcal T=\Pi_n$&$\mathcal T=z\Pi_n$\\\hline
$\vec n=\vec r/r$&$\vec n=-\vec r/r$&-$\vec n=\vec r/r$&$\vec n=\vec r/r$\\
$q_0\ge 0$&$q_0\le 0$&$q_0\ge 0$&$q_0\le 0$\\\hline
$Z=1$&$Z=-1$&$Z=-1$&$Z=1$\\
$\mathcal Q=\frac{1}{2}$&$\mathcal Q=\frac{1}{2}$&$\mathcal Q=-\frac{1}{2}$&$\mathcal Q=-\frac{1}{2}$\\\hline
\includegraphics[scale=0.18]{figures/schemaigel}&\includegraphics[scale=0.18]{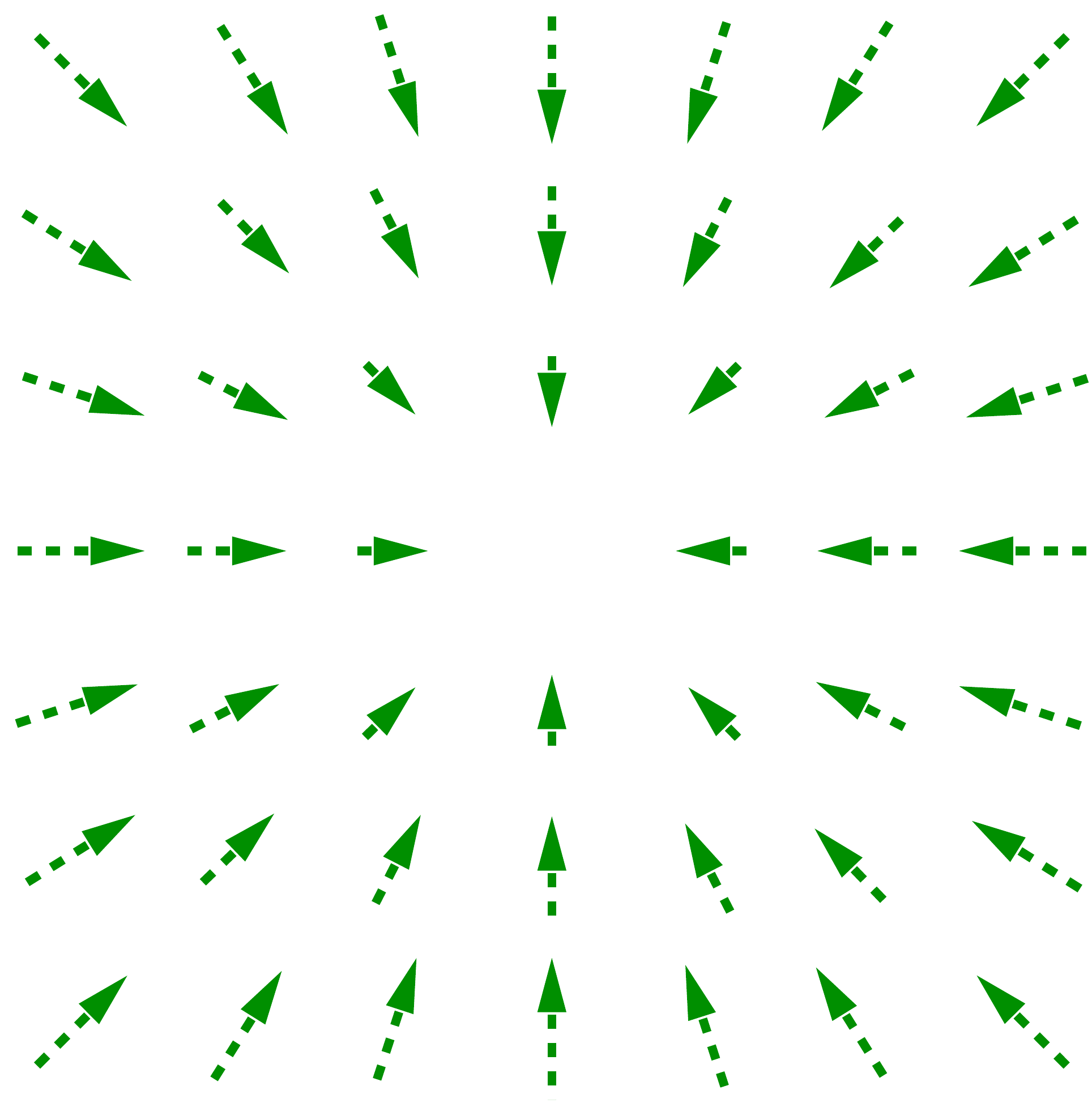}&\includegraphics[scale=0.18]{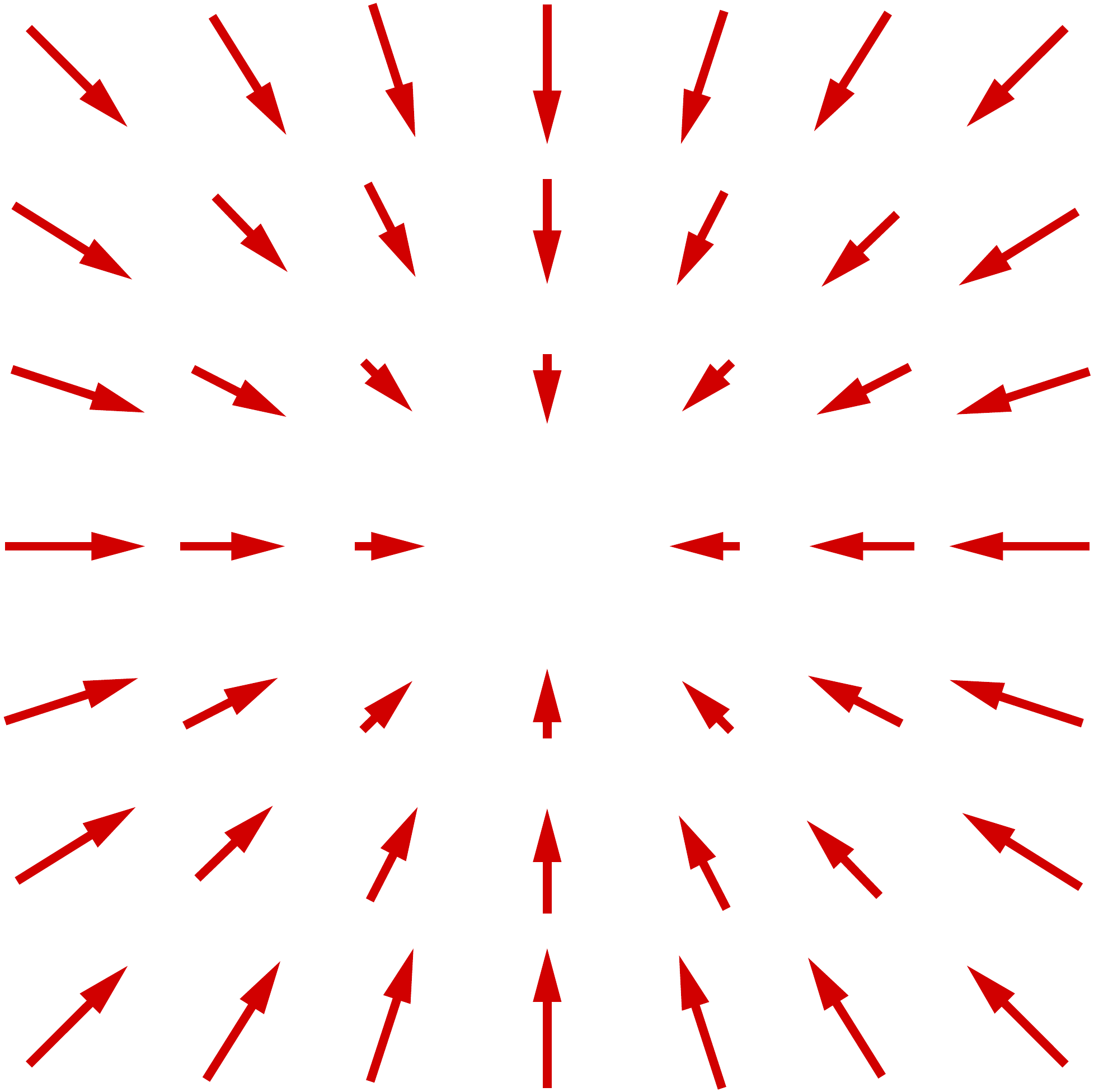}&\includegraphics[scale=0.18]{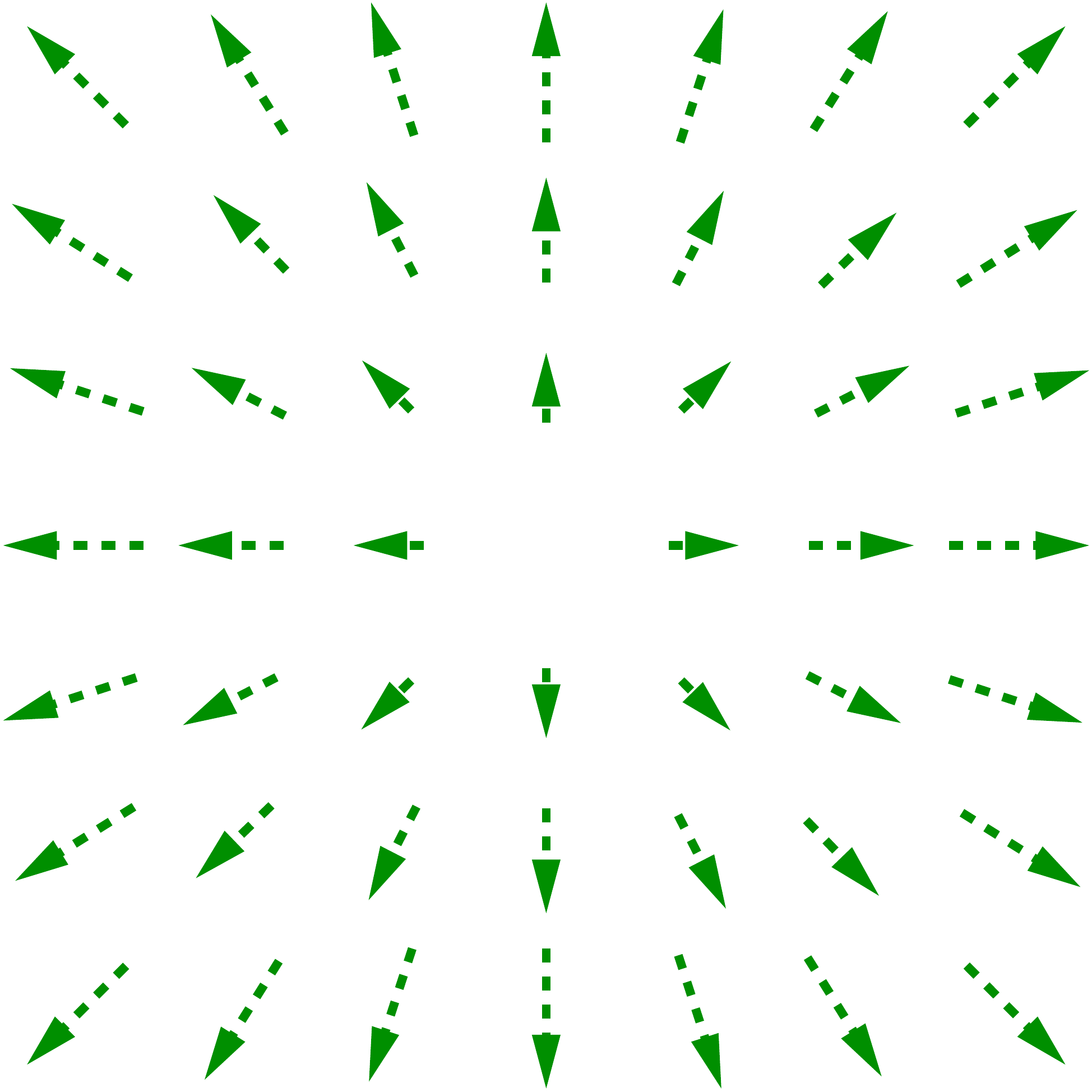}\\\hline
\end{tabular}\caption{The transformations $\mathcal T$ of the hedge-hog configuration in the first column, see Equation~(\ref{RegularIgel}), modify the fields $\vec n$ and $q_0$ and the topological quantum numbers $Z$ and $\mathcal Q$. The diagrams show the imaginary components $\vec q=\vec n\sin\alpha$ of the soliton field, in full red for the hemisphere with $q_0>0$ and in dashed green for $q_0<0$.}
\label{TabSign}\end{center}\end{table}
Configurations with vectors $\vec q$ pointing outwards/inwards are characterised by $Z=+1/-1$, respectively. Crossing the center of the configurations in an arbitrary direction leads to a rotation by $\pm2\pi$. We get a right-handed rotation, $2\pi$, for the configurations with $\mathcal Q=\frac{1}{2}$ and a left-handed rotation, $-2\pi$, for $\mathcal Q=-\frac{1}{2}$. With the sign of $\mathcal Q$ we define an internal chirality $\chi$ and with the absolute value of $\mathcal Q$ the spin quantum number $s$,
\begin{equation}\label{DefChiral}
\mathcal Q=\chi\cdot s\quad\textrm{with}\quad s:=|\mathcal Q|.
\end{equation}
The spin quantum number of two-soliton configurations indicated that $\chi$ can be related to the sign of the magnetic spin quantum number.

Eq.~(\ref{DefChiral}) may possibly give an answer to a question which T.D.Lee posed in a talk at the University of Vienna around 1990: ``Why does mass violate chiral symmetry?''

Motivated by the definition.~(\ref{TopolChar}) we introduce the topologcial current
\begin{equation}\begin{split}\label{DefTopStrom}
k^\mu&:=\frac{\epsilon^{\mu\nu\rho\sigma}}{12\pi^2}
\vec\Gamma_\nu(\vec\Gamma_\rho\times\vec\Gamma_\sigma)
=-\frac{\epsilon^{\mu\nu\rho\sigma}}{24\pi^2}\mathrm{Sp}[
(-\mathrm i\vec\sigma\vec\Gamma_\nu)(-\mathrm i\vec\sigma\vec\Gamma_\rho)
(-\mathrm i\vec\sigma\vec\Gamma_\sigma)]=\\
&\ist{derivativeQ}-\frac{\epsilon^{\mu\nu\rho\sigma}}{24\pi^2}\mathrm{Sp}[
\partial_\nu QQ^\dagger\partial_\rho QQ^\dagger\partial_\sigma QQ^\dagger]
=-\frac{\epsilon^{\mu\nu\rho\sigma}}{24\pi^2}\mathrm{Sp}
\left[Q\partial_\nu Q^\dagger\partial_\rho Q\partial_\sigma Q^\dagger\right]
\end{split}\end{equation}
Regions of $k^\mu\ne 0$ turn out to be identical to regions with $\mathcal L_\mathrm{pot}\ne 0$. $k^\mu$ is obviously a conserved current
\begin{equation}\begin{split}\label{ConsTopCurr}
\partial_\mu k^\mu&\ist{DefTopStrom}-\frac{\epsilon^{\mu\nu\rho\sigma}}{12\pi^2}\mathrm{Sp}\left[
\partial_\mu Q\partial_\nu Q^\dagger\partial_\rho Q\partial_\sigma Q^\dagger\right]=\\
&=-\frac{\epsilon^{\mu\nu\rho\sigma}}{12\pi^2}\mathrm{Sp}\left[\partial_\mu QQ^\dagger
\partial_\nu QQ^\dagger\partial_\rho QQ^\dagger\partial_\sigma QQ^\dagger\right]=0.
\end{split}\end{equation}
If a soliton is leaving the investigated volume the topological charge will change. The conservation of $k^\mu$ has the interesting consequence of vanishing $\vec F^{\mu\nu}\hspace{0.5mm}{^\star}\hspace{-0.8mm}\vec F_{\mu\nu}\propto\vec{\hspace*{0.5mm}\mathbf E}\vec{\hspace*{0.5mm}\mathbf B}$,
\begin{equation}\begin{aligned}\label{EBNull}
0&\ist{ConsTopCurr}\epsilon^{\mu\nu\rho\sigma}\mathrm{Sp}\left[\partial_\mu QQ^\dagger
 \partial_\nu QQ^\dagger\partial_\rho QQ^\dagger\partial_\sigma QQ^\dagger\right]=\\
 &\ist{derivativeQ}\epsilon^{\mu\nu\rho\sigma}\mathrm{Sp}[(\mathrm i\vec\sigma\vec\Gamma_\mu)
 (\mathrm i\vec\sigma\vec\Gamma_\nu)(\mathrm i\vec\sigma\vec\Gamma_\rho)
 (\mathrm i\vec\sigma\vec\Gamma_\sigma)]=\\
 &=-2\,\epsilon^{\mu\nu\rho\sigma}(\vec\Gamma_\mu\times\vec\Gamma_\nu)
 (\vec\Gamma_\rho\times\vec\Gamma_\sigma)
 =-2\epsilon^{\mu\nu\rho\sigma}\vec R_{\mu\nu}\vec R_{\rho\sigma}\propto\vec F^{\mu\nu}\hspace{0.5mm}{^\star}\hspace{-0.8mm}\vec F_{\mu\nu}.
\end{aligned}\end{equation}

The four configurations of Tab.~\ref{TabSign} correspond to the four components of Dirac spinors, spin up $\uparrow$ and down $\downarrow$, electron $e_-$ and positron $e_+$. In the order of Tab.~\ref{TabSign} we can attribute to the configurations the particle types and spin components
\begin{equation}\label{TeilchenSpin}
\overset{\uparrow}{e_-},\;\overset{\uparrow}{e_+},\;
\overset{\downarrow}{e_-},\;\overset{\downarrow}{e_+}
\end{equation}

In relativistic quantum mechanics we can attribute to particles an intrinsic parity and use $\gamma_0$ as parity operator. The Dirac equation has two pairs of solutions of opposite parity. This abstract parity property has an anschauliche realisation in the figures of Tab.~\ref{TabSign} with two pairs of configurations with opposite parity.

%------------------------------------------------------------------------------
\subsubsection{Spin as angular momentum}\label{Sec-Spin}
%------------------------------------------------------------------------------
Spin properties are well-known from quantum mechanics. They are understood as internal properties of particles, characterised by the spin quantum number $s$ taking half integer values as described by the group SU(2) and its representations. Similar, but more general quantisation rules, those of SU(3), are known for the the color of quarks. In distinction to the internal color property in QCD and the Standard Model, spin contributes to the total angular momentum. This property is the origin of one of the enigmas of quantum mechanics, since this fundamental theory does not tell us what is really rotating, especially due to the point-like nature of fundamental particles which is generally assumed. Further, the spin is associated with a contribution to the magnetic moment of particles which in all other circumstances has its origin in the motion of charges.

It is an interesting question whether the present model may lead to some deeper understanding of these riddles. It seems obvious that stable SO(3) field configurations covering half of the SU(2) group manifold follow the quantisation rules of SU(2) for the defining two-dimensional representation and its tensor products.
\begin{figure}[h!]
\hspace*{-11mm}\includegraphics[scale=0.6]{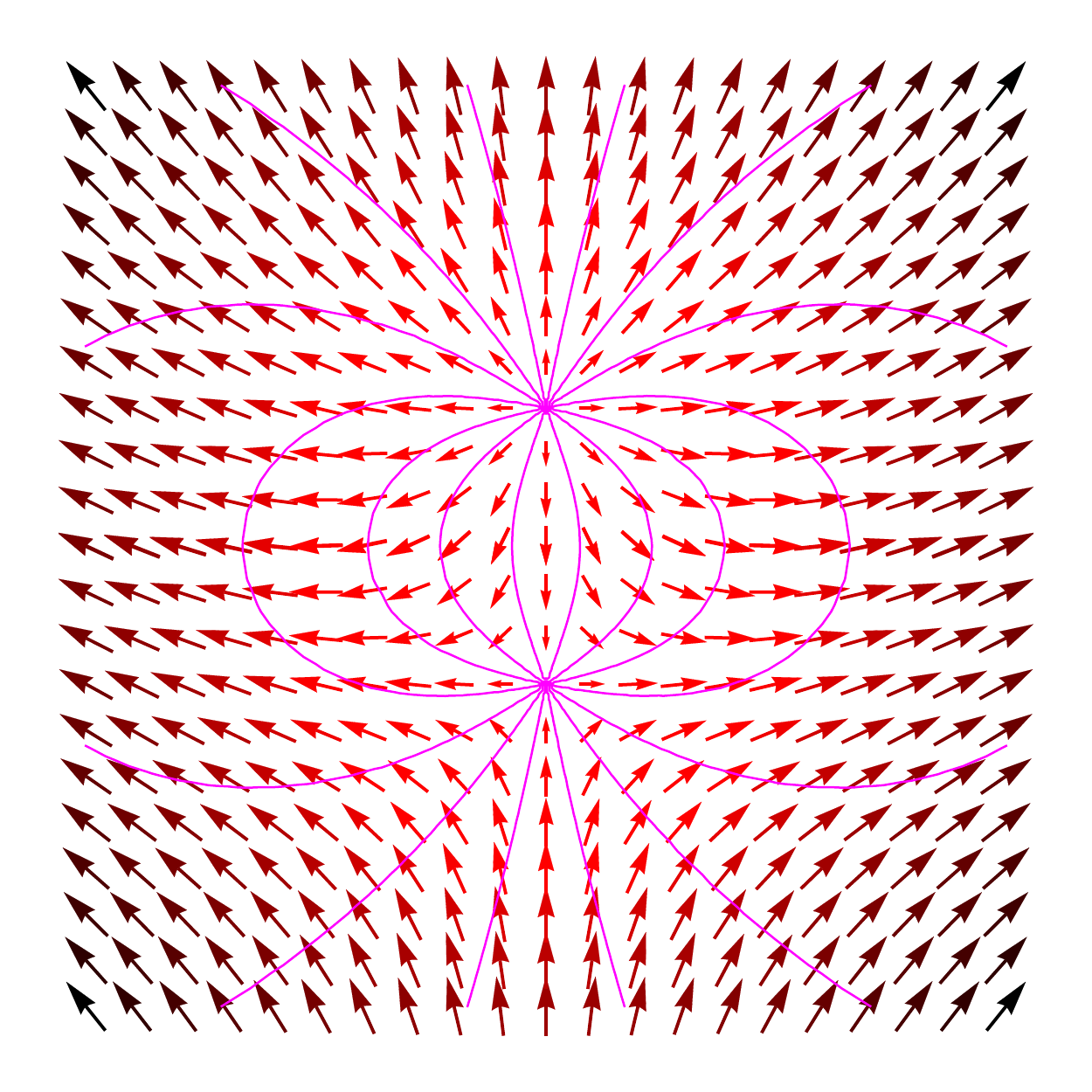}\hspace{5mm}
\includegraphics[scale=0.6]{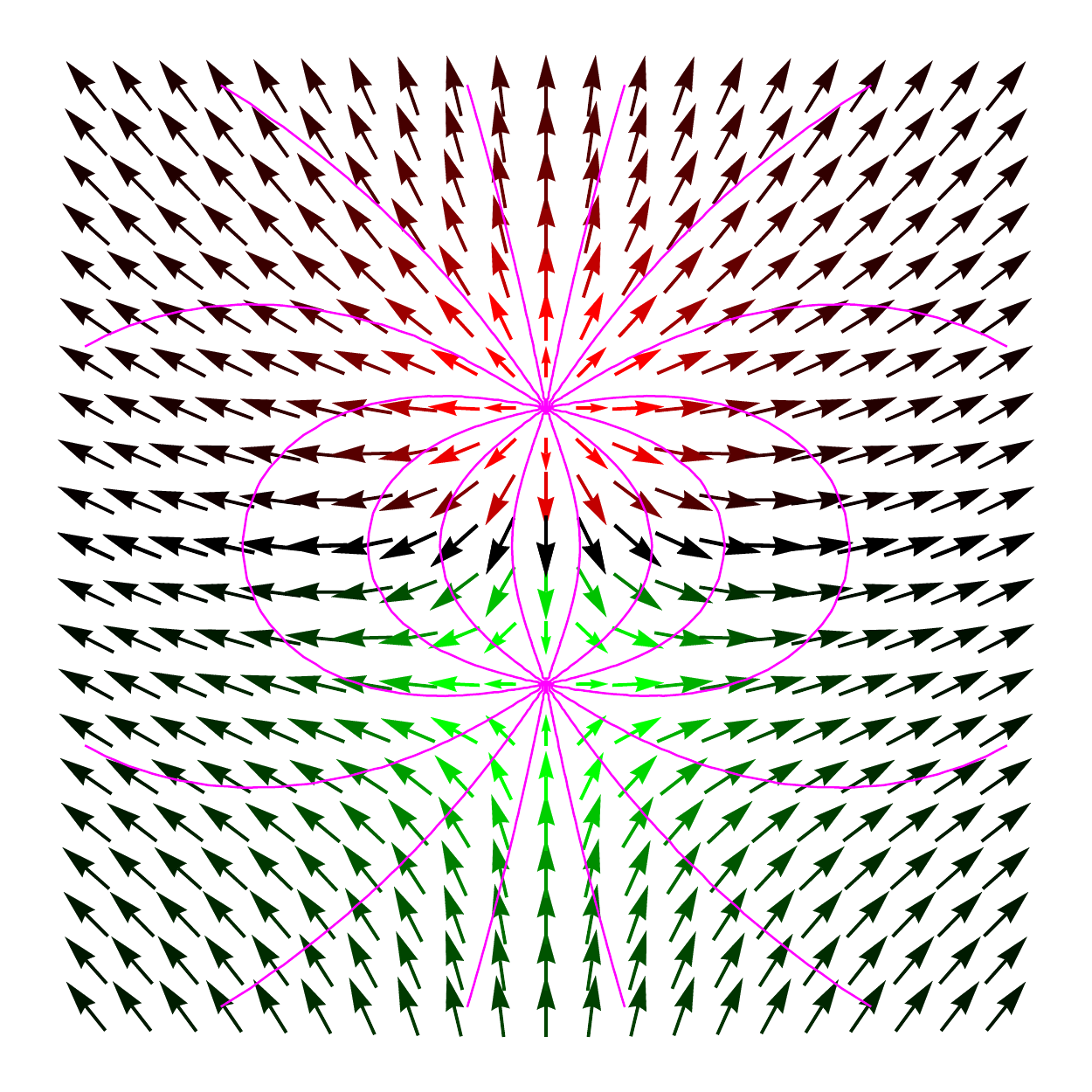}
\caption{Schematic diagrams for the $Q$-field (arrows) and the flux lines (lines) of two opposite unit charges, $Z=Q_\mathrm{el}=0$. The configurations are rotational symmetric around the axis through the two charge centers. The red/green arrows symbolise values of $\vec q=\vec n\sin\alpha$ for postive/negative values of $q_0=\cos\alpha$. For $q_0\to0$ the arrows are getting darker or black. The left configuration belongs to the topological quantum numbers $\mathcal Q=s=0$ and the left one to $\mathcal Q=s=1$. Observe that field lines connect points of constant $\vec n$-field.}
\label{anziehung}
\end{figure}
In fact, we can see that we can combine the four single-particle configurations of Table~\ref{TabSign} to $\binom{4}{3}=10$ topologically different configurations. The charge numbers $Z=\pm1$ combine to $Z=0$ or $\pm2$ and the topological charges $\mathcal Q=\pm1/2$ to $\mathcal Q=0$ or $\mathcal Q=\pm1$. The combination has to be done by appropriate rotations of one of the single-particle configurations. In this way it is possible to merge the two different configurations and to minimise the energy, fixing the centers of the solitons, as will be described in more detail in a forthcoming paper. In the framework of our SO(3)-model we have to respect that configurations differing in the sign of $\mathcal Q$ are physically identical. In Fig.~\ref{anziehung} we show schematic diagrams for two configurations with $Z=0$. They are axially symmetric around the vertical axis. The vectors $\vec q$ of the imaginary parts of the soliton field are plotted. The left diagram with $q_0\ge0$ everywhere and spin quantum number $s=0$ differs slightly from the right diagram for $s=1$, with $q_0\ge0$ in the upper half and $q_0\le0$ in the lower half for this attractive configuration. Comparing the single particle configurations in Table~\ref{TabSign} with Fig.~\ref{anziehung} one can observe that the lower soliton configurations, the antiparticles are rotated by $\pi$ around the vertical symmetry axis. For comparison some electric field lines are drawn. They agree with lines $\vec n$=const. This agreement can be used to find an appropriate initial configuration for a minimisation process.

\begin{figure}[h!]
\centering
\includegraphics*[scale=0.2]{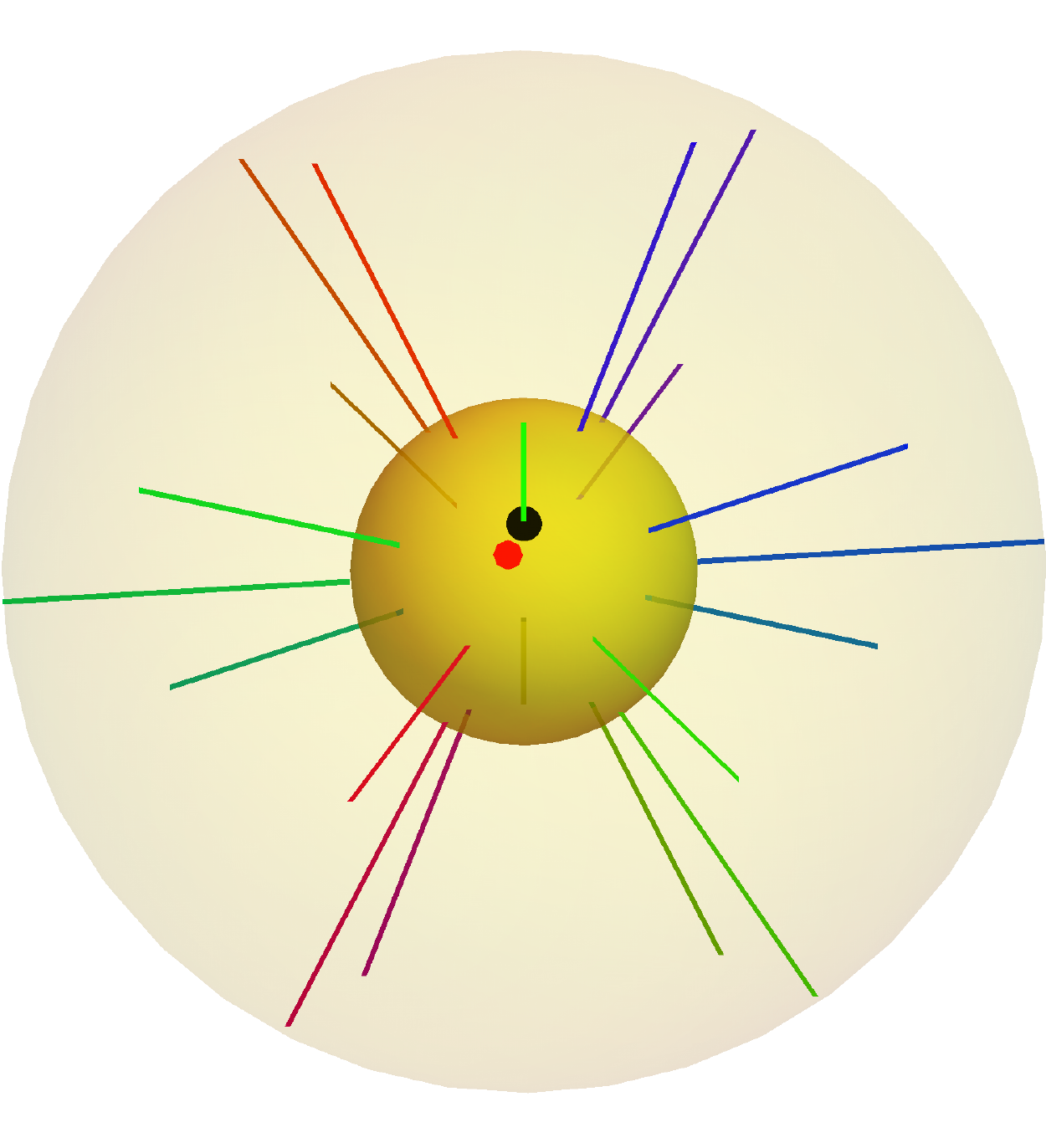}
\includegraphics*[scale=0.2]{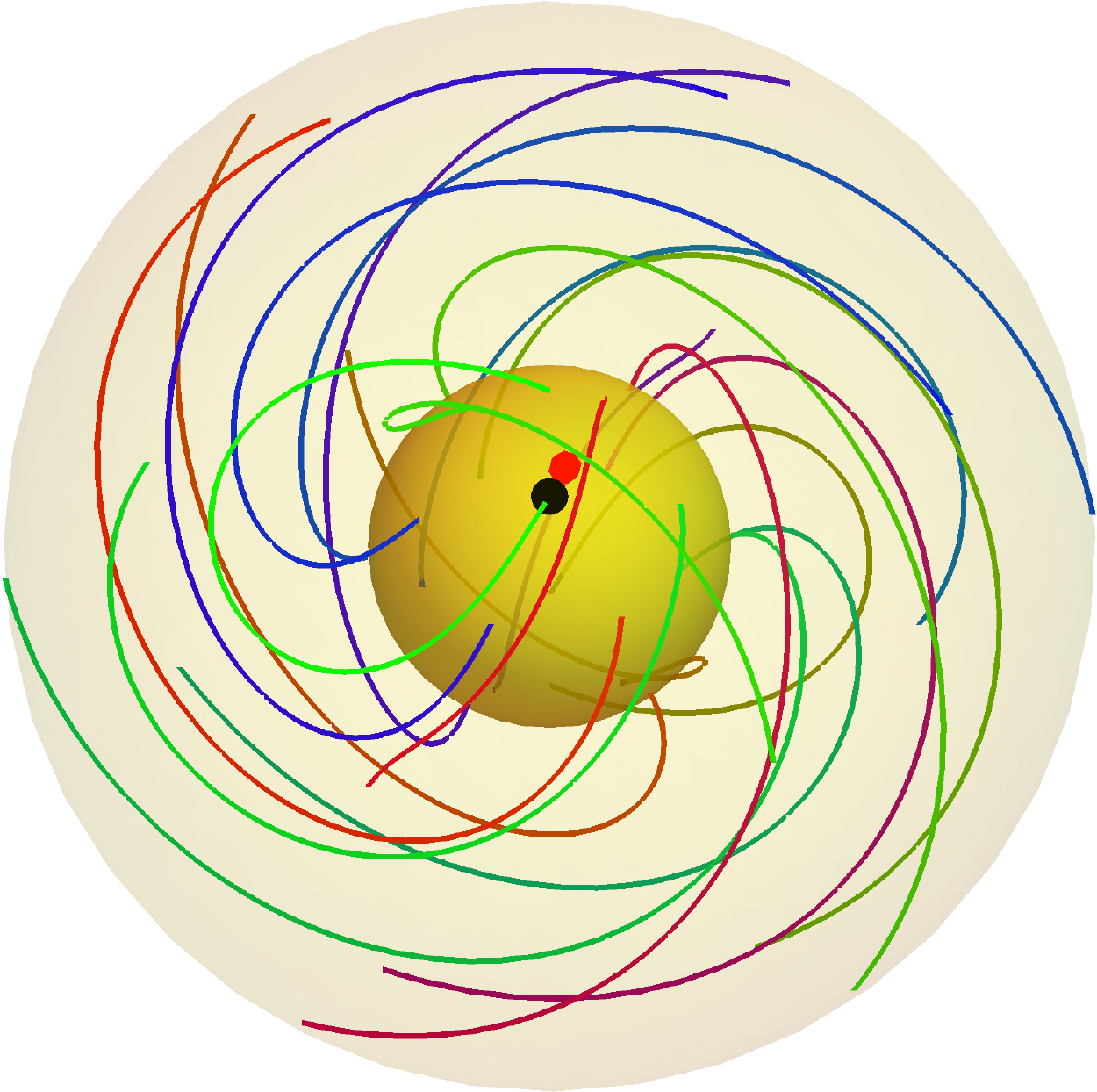}
\includegraphics*[scale=0.2]{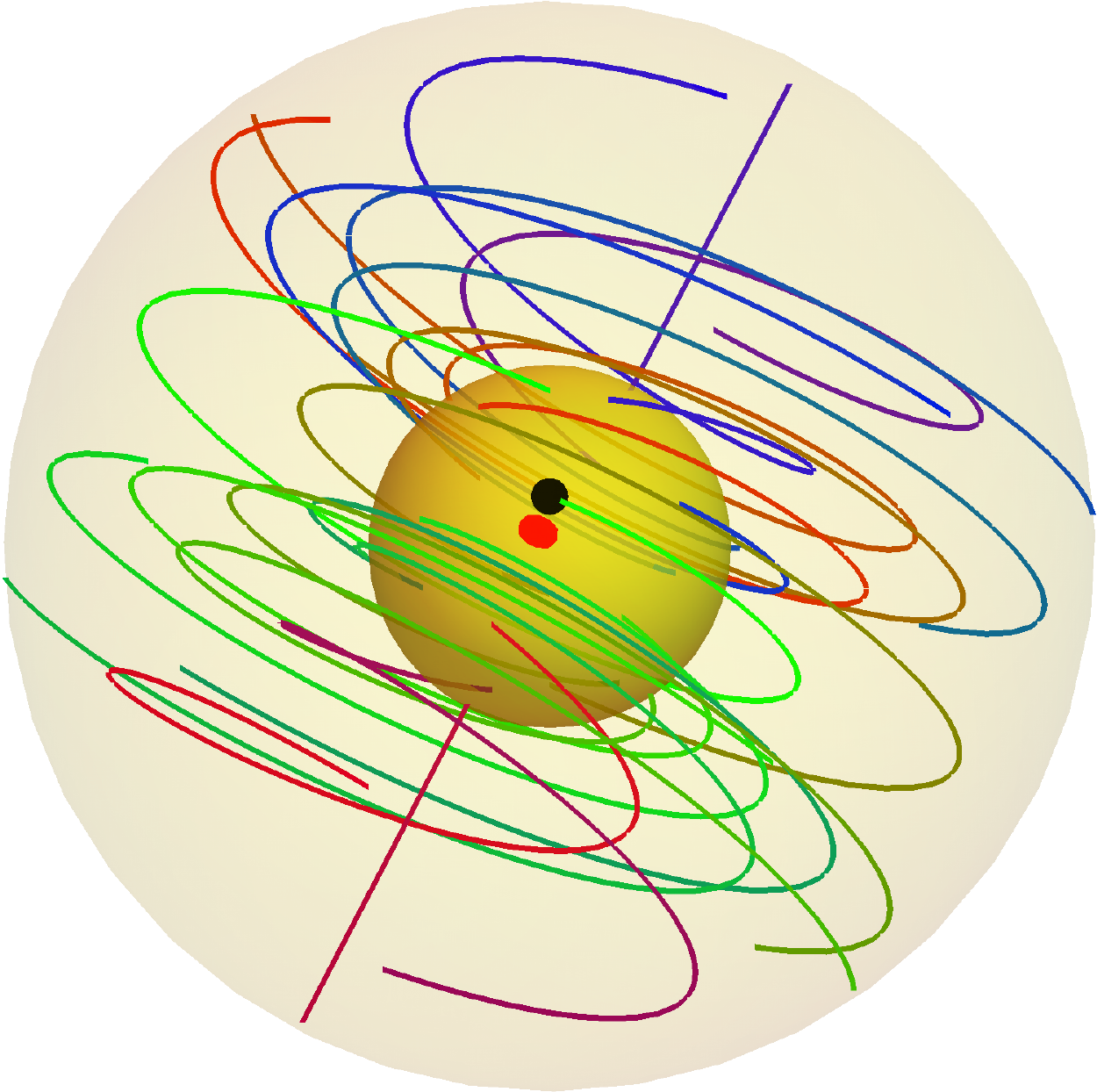}
\includegraphics*[scale=0.2]{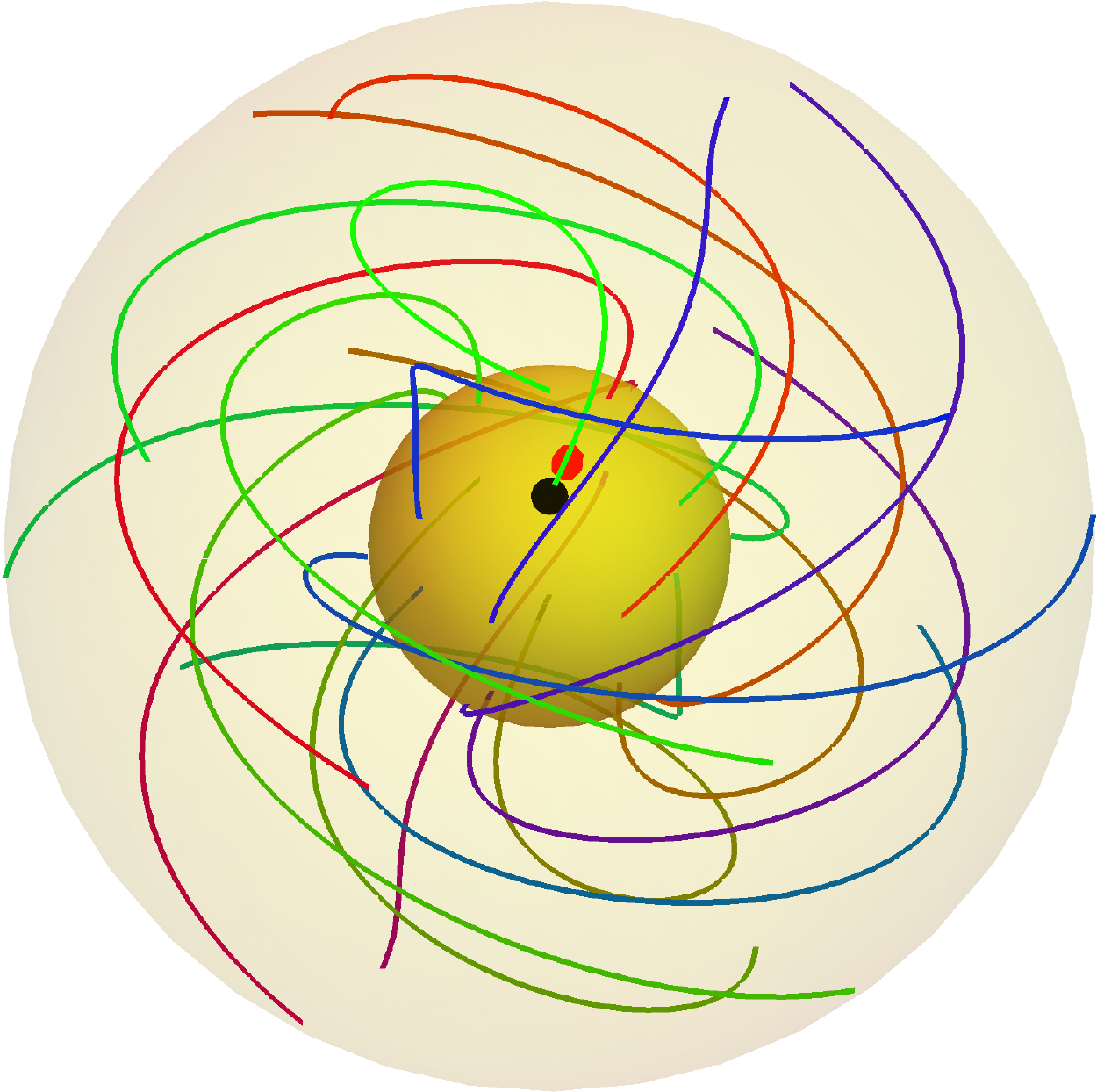}
\includegraphics*[scale=0.2]{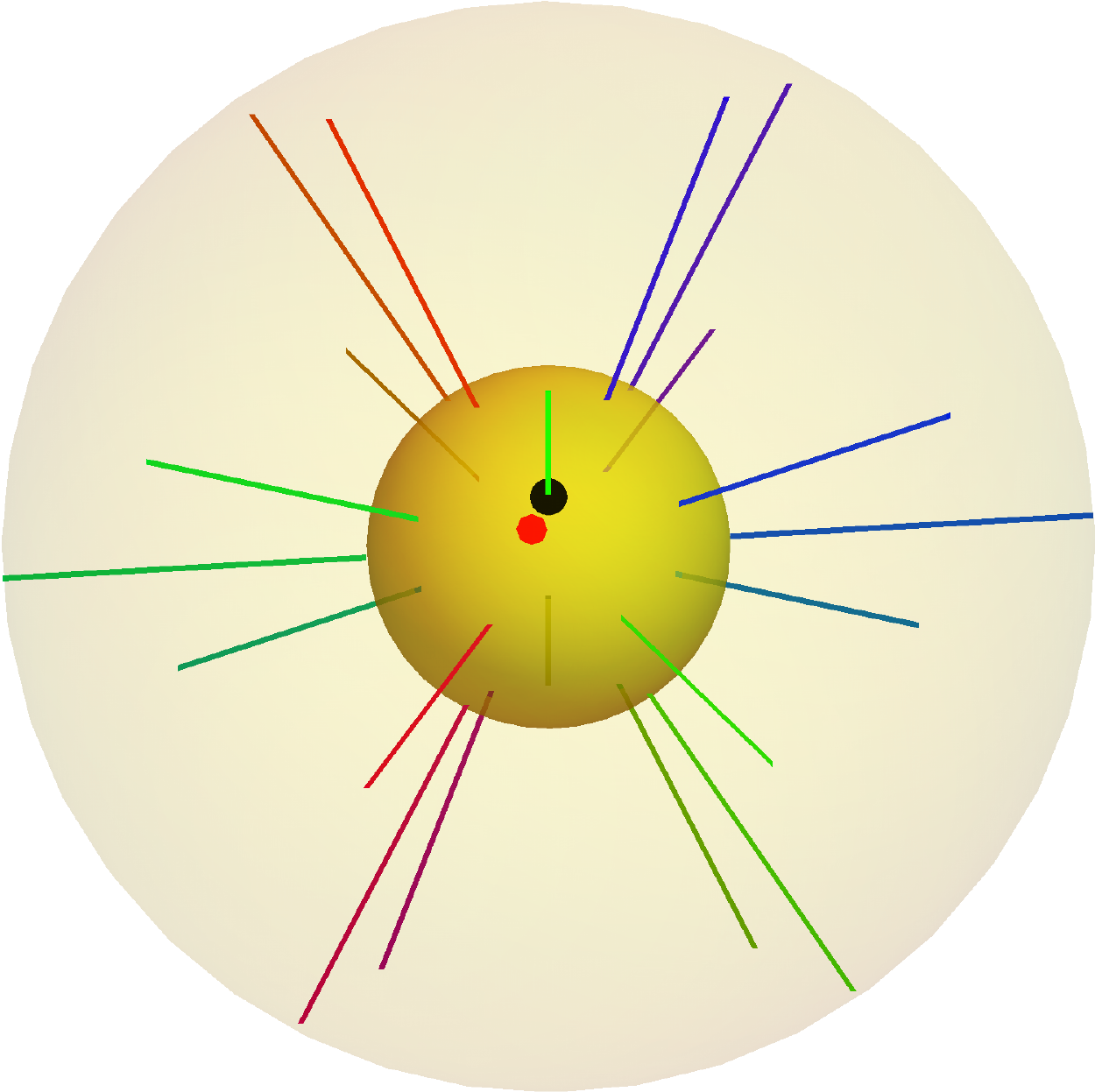}
\caption{The rotating inner sphere is connected by an arbitrary number of wires to the outer sphere. Rigid spheres of increasing radius with fixed piercings rotate around varying axes. The inner sphere performs a $4\pi$ rotation around the axis indicated by the black dot. The red dot shows the status of the positive rotation in the sequence $0,\pi,2\pi,3\pi,4\pi$. The set of these rotations corresponds to a big circle on $\mathbb S^3$. For spheres from inside to outside, these sets build smaller and smaller circles contracting at the outer sphere to no rotation.}
\label{verwirrt}
\end{figure}

A very famous property of spin-1/2 particles is their $4\pi$-periodicity under rotations. It is well-known in mathematics, even if not all mathematicians and physicists know about it: Objects which are connected with a fixed surrounding feel this $4\pi$-periodicity. This can easily be shown in a nice tiny experiment with an around decimetre sized sphere which is connected by stretchable rubber bands with a surrounding 3-dimensional frame. After a $2\pi$-rotation these wires are terribly entangled and even more after a $4\pi$-rotation, but then the wires can be nicely disentangled without further moving the central sphere by some skilful movements of the wires. More regularly this is shown in Fig.~\ref{verwirrt} where the rotation angle of the inner sphere increases from left to right from 0 to $4\pi$, as indicated by the red dot rotating around the black dot where the rotational axis pierces. The other shells and their piercings perform rigid rotations around varying axes. The sets of these rotations build lassos on $\mathbb S^3$ contracting from inside to outside from a big circle to a point. 
Exactly this $4\pi$-periodicity is shared by the soliton configurations in Table~\ref{TabSign} which are connected to the sphere at spatial infinity by imaginary rubber bands of constant $\vec n$-vectors. This may be a hint that spin-1/2 particles may be extended objects, connected to the surrounding by some fields, e.g. electrons by their electric field lines.

\begin{figure}[h!]
\hspace*{-12mm}
\includegraphics[scale=0.6]{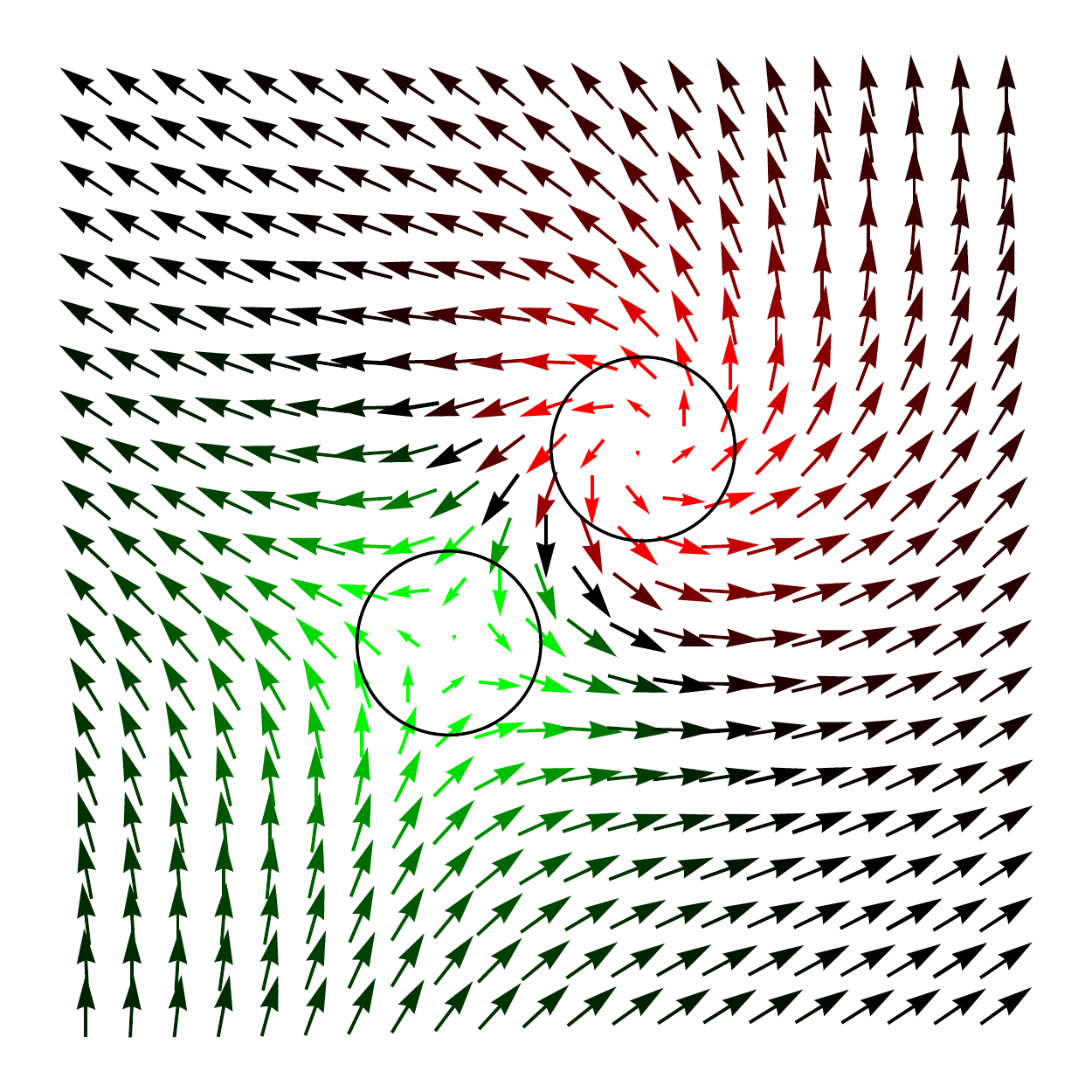}\hspace{5mm}
\includegraphics[scale=0.6]{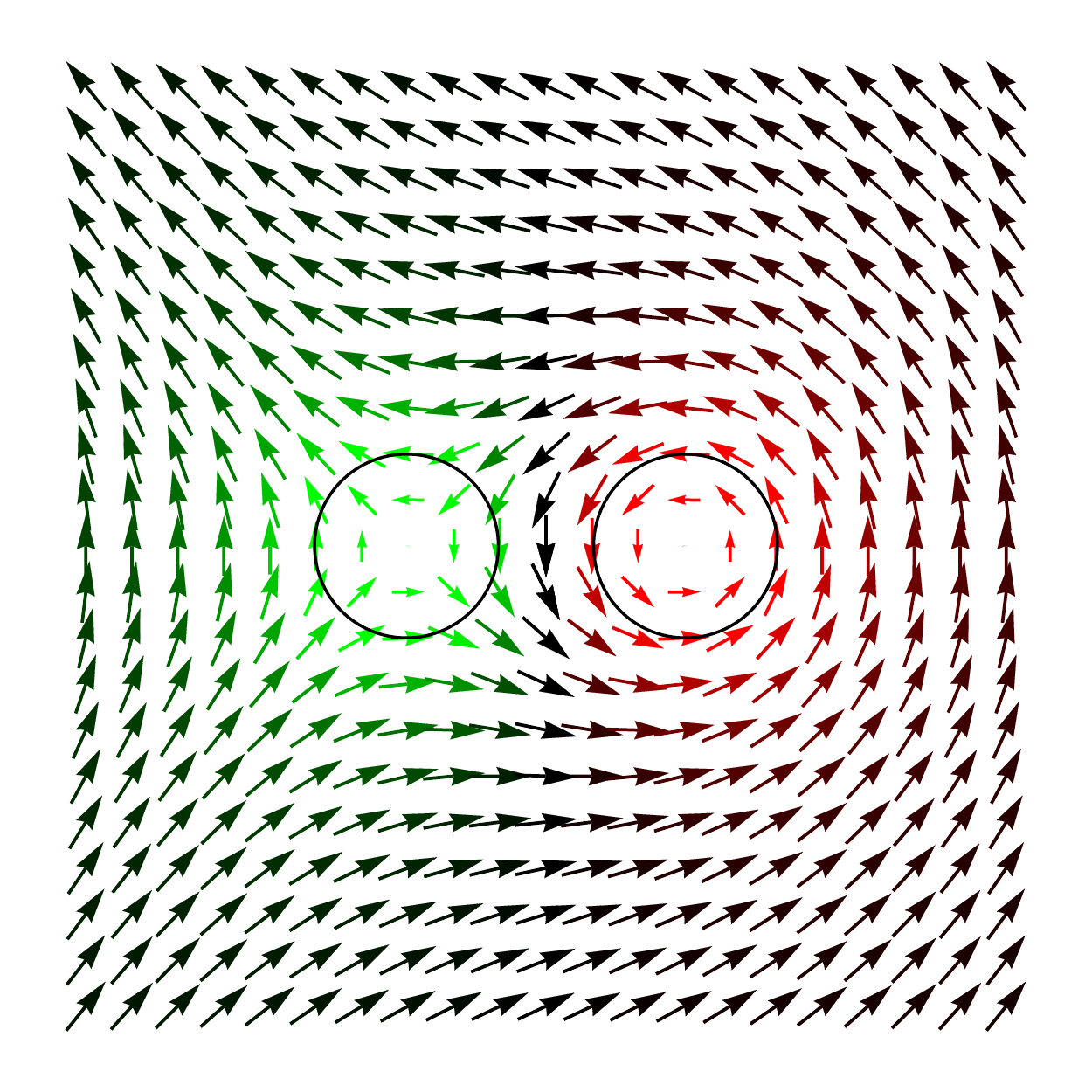}
\caption{Two schematic diagrams for a rotating dipole which differ by rotations of 45$^\circ$ and 90$^\circ$ from Fig.~\ref{anziehung}b. The imaginary part of the SU(2)-field are indicated by arrows. The visibility of the positions of the two charges is enhanced by surrounding circles. Red/green/black arrows indicate $q_0>0/q_0<0/q_0=0$.}
\label{rotating}
\end{figure}

In the first moment it may be disturbing to compare the discussed stable soliton solutions with electrons, since these stable solutions have no internal angular momentum and no magnetic moment, they are just static and spherically symmetric. But, in this soliton model spin angular momentum appears as a consequence of orbital motion. The fact that orbital motion leads to internal rotation can easily been shown in an animation of such a rotation. It is a pity that it is difficult to show this behaviour in an article. To give an impression, two of such diagrams are shown in Fig.~\ref{rotating} and can be compared to Fig.~\ref{anziehung}b.
The internal rotation of the two solitons can be observed in the region of the two circles. It is caused by the broken symmetry of the vacuum with $Q=-\mathrm i\vec\sigma(0,0,1)$, approached by the soliton field at large distances for integer topological charges $\mathcal Q$. It forces the center of the solitons to rotate with the same frequency in the opposite direction. The two different colors indicate $q_0>0$ and $q_0<0$. As well-known from quantum mechanics the internal angular momentum has a fixed relation to the orbital angular momentum, the speed of the internal rotation can not be enlarged or diminished. It will be interesting to compare the internal angular momentum to the orbital angular momentum in analytical or numerical calculations. There is an energy difference between the configurations for $s=0$ and $s=1$ which should contribute to the spin-orbit interaction.

Just like spin angular momentum, the magnetic moment is in this model a dynamical property, a reaction to external magnetic fields. Such magnetic fields with sources in external moving charges should be encoded in a periodic time dependence of the soliton field in the ``vacuum''. These external fields may lead to internal rotation of solitons and to orbital motion.

Dipole configurations of soliton pairs with opposite charge number are unstable due to Coulomb attraction. They can be stabilised in numerical simulations, fixing the position of the soliton centers. In several diploma theses \cite{Wabnig2001,Resch2011,Theuerkauf2016} it turned out that the numerical minimisation of dipole configurations suffers from a numerical instability making it difficult to avoid the annihilation of dipoles with small separations. It is caused by an instability of single monopole configurations due to numerical errors. This problem was solved in~\cite{Anmasser2021} and the accuracy of the solution tested for single monopoles in~\cite{Anmasser:2021ubc}.

% xxx belt trick

%------------------------------------------------------------------------------
\section{General equations of motion}\label{Sec-EOM}
%------------------------------------------------------------------------------
Eq.~(\ref{nlDE}) is the differential equation which results from the Lagrangian~(\ref{Lagr4D}) for a static spherical symmetric configuration. For a general field configuration we get the momentum density~(\ref{VerallgImp}) and the Euler-Lagrange equation~(\ref{BGl1}) by a variation of the soliton field $Q$
\begin{equation}\begin{split}\label{varyQ}
Q\rightarrow Q^\prime &:=\mathrm e^{-\mathrm i\vec\sigma\vec\zeta} Q\ist{unitquaternions}
\left[1-\mathrm i\vec\sigma\vec\zeta\right](q_0-\mathrm i\vec q\vec\sigma)=\\
&=q_0-\vec\zeta\vec q-\mathrm i\vec\sigma\left(\vec q+q_0\vec\zeta+\vec\zeta\times\vec q\right).
\end{split}\end{equation}
We get up to first order in $\zeta$
\begin{align}\label{deltaq0}
&\delta q_0=-\vec\zeta\vec q\\\label{DeltaGamma}
&\delta\vec\Gamma_\mu\ist{derivativeQ}\frac{\mathrm i}{2}\mathrm{Sp}
\{[\partial_\mu(Q+\delta Q)(Q^\dagger+\delta Q^\dagger)
-\partial_\mu Q Q^\dagger]\vec\sigma\}
=\partial_\mu\vec\zeta-2\vec\Gamma_\mu\times\vec\zeta\\
&\delta\vec R_{\mu\nu}\ist{RSU2}\partial_\mu\vec\zeta\times\vec\Gamma_\nu
+2\vec\Gamma_\mu(\vec\zeta\vec\Gamma_\nu)
-\partial_\nu\vec\zeta\times\vec\Gamma_\mu
-2\vec\Gamma_\nu(\vec\zeta\vec\Gamma_\mu).\label{dRmn}
\end{align}
From $\vec\Gamma_\nu\vec R^{\mu\nu}\ist{RSU2}0$ follows for the square of the curvature
\begin{equation}\label{deltaRquadrat}
\delta(\vec R_{\mu\nu}\vec R^{\mu\nu})
=2\delta\vec R_{\mu\nu}\vec R^{\mu\nu}
\ist{dRmn}4(\partial_\mu\vec\zeta\times\vec\Gamma_\nu)\vec R^{\mu\nu}
=4\partial_\mu\vec\zeta(\vec\Gamma_\nu\times\vec R^{\mu\nu}).
\end{equation}
From the variation of the Lagrange density~(\ref{Lagr4D})
\begin{equation}\label{VarLagrange}
-\frac{4\pi}{\alpha_f\hbar c_0}\delta\mathcal L\ist{Lagr4D}
\delta\left(\frac{1}{4}\,\vec R_{\mu\nu}\vec R^{\mu\nu}+\Lambda(q_0)\right)
\iist{deltaRquadrat}{deltaq0}-(\vec\zeta\vec q)\partial_{q_0}\Lambda 
+\partial_\mu\vec\zeta(\vec\Gamma_\nu\times\vec R^{\mu\nu})
\end{equation}
we read the momentum density
\begin{equation}\label{VerallgImp}
\vec\pi^\mu:=\frac{\partial\mathcal L}{\partial_\mu\vec\zeta}
\ist{VarLagrange}-\frac{\alpha_f\hbar c_0}{4\pi}\vec\Gamma_\nu\times\vec R^{\mu\nu}
\ist{Lagr4D}\frac{\partial\mathcal L}{\partial\vec\Gamma_\mu}.
\end{equation}
The last expression for $\vec\pi^\mu$ shows that $\vec\Gamma_\mu$ can be regarded as generalised velocity. Further we get the general equation of motion
\begin{equation}\label{BGl1}
\partial_\mu[\vec\Gamma_\nu\times\vec R^{\mu\nu}]
+\vec q\,\frac{\mathrm d\Lambda}{\mathrm d q_0}\ist{deltaRquadrat}0\quad
\Leftrightarrow\quad\partial_\mu\vec\pi^\mu
+\vec q\,\frac{\mathrm d\mathcal L_\mathrm{pot}}{\mathrm d q_0}\ist{deltaRquadrat}0
\end{equation}
describing that the variation of the potential is the source of the momentum current. With $\vec R_{\mu\nu}\iist{RSU2}{RAllg}\frac{1}{2}(\partial_\mu\vec\Gamma_\nu-\partial_\nu\vec\Gamma_\mu)$ and $\vec R_{\mu\nu}\times\vec R^{\mu\nu}=0$ we can get another form of the equation of motion
\begin{equation}\label{BGl2}
\vec\Gamma_\nu\times\partial_\mu\vec R^{\mu\nu}+\vec q\,
\frac{\mathrm d\Lambda}{\mathrm dq_0}=0.
\end{equation}
A consequence of the equation of motion is
\begin{equation}\label{GradLpot}
\vec\Gamma_\nu\partial_\mu\vec\pi^\mu\ist{BGl1}
-\vec\Gamma_\nu\vec q\,\frac{\mathrm d\mathcal L_\mathrm{pot}}{\mathrm dq_0}
\ist{ConCof}-\partial_\nu\alpha\sin\alpha\,
\frac{\mathrm d\mathcal L_\mathrm{pot}}{\mathrm dq_0}\ist{nalpha}
\partial_\nu q_0\,\frac{\mathrm d\mathcal L_\mathrm{pot}}{\mathrm dq_0}
=\partial_\nu\mathcal L_\mathrm{pot}.
\end{equation}

In Sect.~\ref{Sec-Edyn} we will separate the system in point-like sources and their fields. This should allow some checks whether these equations describe the dynamics of these solitons and their interactions with electromagnetic fields correctly.

Using the generalised velocity $\vec\Gamma_\mu$ it is easy to derive the energy momentum tensor
\begin{equation}\begin{aligned}\label{DefEMT}
{\Theta^\mu}_\nu(x)&:=\frac{\partial\mathcal L(x)}{\partial\vec\Gamma_\mu}\,
\vec\Gamma_\nu-\mathcal L(x)\,\delta^\mu_\nu
\ist{VerallgImp}\vec\pi^\mu\vec\Gamma_\nu-\mathcal L(x)\,\delta^\mu_\nu=\\
&\iist{VerallgImp}{RSU2}-\frac{\alpha_f\hbar c_0}{4\pi}
\left\{\left(\vec\Gamma_\nu\times\vec\Gamma_\sigma\right)
\left(\vec\Gamma^\mu\times\vec\Gamma^\sigma\right)\right\}-
\mathcal L(x)\,\delta^\mu_\nu .
\end{aligned}\end{equation}
In distinction to the energy momentum tensor in Maxwell's theory, it is automatically symmetric. Its trace is determined by the potential energy
\begin{equation}\label{SpurEnMomTens}
  {\Theta^\mu}_\mu\iist{DefEMT}{Lagr4D}-4\mathcal L_\mathrm{pot}\ist{Lagr4D}
  \frac{\alpha_f\hbar c_0}{4\pi}\Lambda(q_0).
\end{equation}

The force density is given by the divergence of the energy momentum tensor 
\begin{equation}\begin{aligned}\label{Kraftdichte}
f_\nu&:=\partial_\mu\Theta^\mu_{\;\nu}\ist{DefEMT}
\vec\Gamma_\nu\partial_\mu\vec\pi^\mu+\underbrace{
\partial_\mu\vec\Gamma_\nu\vec\pi^\mu -\partial_\nu\vec\Gamma_\mu\vec\pi^\mu
}_{2\vec R_{\mu\nu}\vec\pi_\mu}-\partial_\nu\mathcal L_\mathrm{pot}=\\
&\ist{RSU2}\vec\Gamma_\nu\partial_\mu\vec\pi^\mu+2\vec R_{\nu\mu}\vec\pi^\mu
-\partial_\nu\mathcal L_\mathrm{pot}\ist{GradLpot}2\vec R_{\nu\mu}\vec\pi^\mu
\iist{VerallgImp}{RSU2}0,
\end{aligned}\end{equation}
where we used $\vec R_{\nu\mu}\vec\pi^\mu=0$. It is finally not astonishing that the force density vanishes in a classical closed system.

To identify the forces between particles and their fields one has to separate artificially particles from fields. This will be done in the next section.

%------------------------------------------------------------------------------
\section{Electrodynamic limit}\label{Sec-Edyn}
%------------------------------------------------------------------------------
In the limit $r_0\to0$ the energy of solitons approaches infinity, we dubbed it the electrodynamic limit~\cite{faber:2002nw}. It suffers from the problem of infinities of the selfenergy of point-like charges, well-known from Maxwell's electrodynamics. Within these models, the only known solution is Kramers suggestion to subtract compensating infinities. In this limit, we do not care of these infinities and arrive at the Wu-Yang formulation of Dirac monopoles~\cite{wu:1975vq,wu:1976qk}
\begin{equation}\label{Qlimit}
Q(x):=-\mathrm i\vec\sigma\vec n(x),\quad\vec n^2:=1,\quad\vec q:=\vec n,\quad
q_0:=\cos\alpha=0,\quad\alpha:=\frac{\pi}{2},
\end{equation}
with values of the soliton field $\vec n(x)$ on $\mathbb S^2$. We use the singularities of the $\vec n$-field at given time $t$ to identify the position of solitons. Since these singularities have no chirality there are only two types of singularities, sources (particles) and sinks (antiparticles) of the $\vec n$-field, with different charge number $Z=\pm1$ and with singularities from the homotopy class $\vec n(\vec r)=Z\frac{\vec r}{r}$.

%------------------------------------------------------------------------------
\subsection{Comparison to Maxwell equations}\label{Sec-EdynMaxwell}
%------------------------------------------------------------------------------
The singular lines of the sources are related to the properties of the connection and the curvature field
\begin{equation}\label{ConWu}
\vec\Gamma_\mu(x)\iist{ConCof}{Qlimit}\vec n(x)\times\partial_\mu\vec n(x),
\end{equation}
\begin{equation}\label{CurvWu}
\vec R_{\mu\nu}(x)\iist{RSU2}{ConWu}
\partial_\mu\vec n(x)\times\partial_\nu\vec n(x).
\end{equation}

The modulus $R_{\mu\nu}:=\vec R_{\mu\nu}\vec n$ defines a dual abelian field strength tensor which reads in SI units
\begin{eqnarray}\label{AbelianFS}
{\hspace{1mm}^\star}\hspace{-1mm}f_{\mu\nu}&
=:&-\frac{e_0}{4\pi\varepsilon_0 c_0}\vec R_{\mu\nu}\vec n
\ist{CurvWu}-\frac{e_0}{4\pi\varepsilon_0 c_0}\,\vec n
(\partial_\mu\vec n\times\partial_\nu\vec n)
\end{eqnarray}
In this limit the Lagrange density reduces to 
\begin{equation}\label{EDLagrangian}
\mathcal L_\mathrm{ED}\ist{Lagr4D}-\frac{\alpha_f\hbar c_0}{4\pi}\frac{1}{4}\,
(\vec n\vec R_{\mu\nu})(\vec n\vec R^{\mu\nu})
\ist{AbelianFS}-\frac{1}{4\mu_0}{\hspace{0.5mm}^\star}\hspace{-0.5mm}
f_{\mu\nu}(x){\hspace{0.5mm}^\star}\hspace{-0.5mm}f^{\mu\nu}(x).
\end{equation}
Up to the sign it has the same form as the Lagrangian of Maxwell's electrodynamics. In order to compare the dynamics resulting from this Lagrangian to Maxwell equations we relate the world-lines of the singularities with the properties of the $\vec n$-field. The singularities are located in infinitesimal space-like three dimensional volumes, the centers of hedgehog configurations of the $\vec n$-field. We describe them by $N$ time-like world-lines of singularities
\begin{equation}\label{worldline}
X_i^\mu(\tau)\qquad i=1,\cdots,N,
\end{equation}
evolving in time with velocities smaller than $c_0$  from $t=-\infty$ to $t=+\infty$, or meeting with lines of opposite charge, where particle pairs are created or annihilated. This allows to define for each $\vec n$-field configuration a singular vector current along the above world-lines
\begin{equation}\label{divFluss}
\sum_{i=1}^NZ_i\int\mathrm d\tau\,\frac{\mathrm dX_i^\kappa(\tau)}{\mathrm d\tau}
\,\,\delta^4(x-X_i(\tau)):=\frac{1}{8\pi}\varepsilon^{\kappa\lambda\mu\nu}\,
\partial_\lambda\{\vec n(\partial_\mu\vec n\times\partial_\nu\vec n)\}.
\end{equation}
For a single soliton resting at the origin with $Z=1,\tau=t,X^\kappa(\tau)=c_0t\delta_0^\kappa$ both sides reduce to a three-dimensional delta-function at the origin $\delta^3(\vec x)\delta_0^\kappa$ for the time-component of the singular vector current. Multiplying with $-e_0c$ we get the current density $j^\kappa$ of point-like charges $-e_0Z$. A further multiplication by $\mu_0$ leads to
\begin{equation}\label{EdynGrenJ}
\mu_0j^\kappa\ist{divFluss}-\partial_\lambda\underbrace{\frac{e_0 c_0\mu_0}{4\pi}
\frac{1}{2}\epsilon^{\kappa\lambda\mu\nu}\,\vec n[
\partial_\mu\vec n\times\partial_\nu\vec n]}_{f^{\kappa\lambda}}\ist{AbelianFS}
\partial_\lambda f^{\lambda\kappa}
\end{equation}
and therefore to the inhomogeneous Maxwell equations
\begin{equation}\label{inhmax}
\partial_\mu f^{\mu\nu}=\mu_0 j^\nu
\quad\Leftrightarrow\quad
\begin{cases}&\nabla\mathbf E=\frac{\rho}{\varepsilon_0},\\
&\frac{1}{\mu_0}\nabla\times\mathbf B-\varepsilon_0\partial_t\mathbf E
=\mathbf j.\end{cases}
\end{equation}
These equations are a result of the separation of the soliton field $Q(x)$ in singular point-like charges and the surrounding $\vec n$-field describing electromagnetic fields.

The second set of equation follows from the general equations of motion (\ref{BGl1})
\begin{equation}\begin{aligned}\label{BewGlEDynLim}
0\ist{BGl1}&\partial_\mu[\vec\Gamma_\nu\times\vec R^{\mu\nu}]
\ist{ConWu}\partial_\mu[(\vec n\times\partial_\nu\vec n)\times
  (\partial^\mu\vec n\times\partial^\nu\vec n)]=\\
 =&\partial_\mu\{\partial_\nu\vec n
  [\vec n(\partial^\mu\vec n\times\partial^\nu\vec n)]\}
 =\partial_\nu\vec n\,
  \partial_\mu[\vec n(\partial^\mu\vec n\times\partial^\nu\vec n)].
\end{aligned}\end{equation}
This is a condition for magnetic currents
\begin{equation}\label{magcur}
\hspace{-2mm}g^\nu:=-\frac{e_0}{4\pi\varepsilon_0}\partial_\mu
[\vec n(\partial^\mu\vec n\times\partial^\nu\vec n)]\ist{AbelianFS}
c_0\,\partial_\mu\hspace{0.2mm}{^\star}\hspace{-0.2mm}f^{\mu\nu}
\;\Leftrightarrow\;\begin{cases}&\rho_\mathrm{mag}=\nabla\mathbf B,\\
&\mathbf g=-\nabla\times\mathbf E-\partial_t\mathbf B.\end{cases}
\end{equation}
We conclude, that in this dual formulation we have soft Dirac monopoles and conserved magnetic currents
\begin{equation}\label{conscurr}
\partial_\nu g^\nu\ist{magcur}0
\end{equation}
which are not quantised. Since $\vec n$ has only two dofs there are only two independent equations in Eq.~(\ref{BewGlEDynLim}). Extending $\partial_\nu\vec n$ by appropriate factors to the dual field strength tensor we get the equations
\begin{equation}\label{obsEOM}
\hspace{0.8mm}{^\star}\hspace{-0.8mm}f_{\mu\nu} g^\nu\iist{BewGlEDynLim}{magcur}0
\quad\Leftrightarrow\quad\begin{cases}
&\mathbf B\,\mathbf g=0,\\
&c_0^2\mathbf B\,\rho_{\rm mag}=\mathbf g\times\mathbf E.\end{cases}
\end{equation}
The spatial components of these equations we read as dual transformations of a vanishing Lorentz force on magnetic currents. The appearance of these currents is a consequence of the topological restrictions to the $\vec n$-field. This is the price we have to pay for the quantisation of electric charges on a classical basis. Solutions of the (linear) Maxwell equations are also solutions of the equations of motion~(\ref{inhmax}) and(\ref{obsEOM}). But the two dofs of the $\vec n$-field are not sufficient to fulfill the homogeneous Maxwell equations exactly and lead to small deviations, to small non-linearities, to magnetic currents. These deviations should remain small for field configurations of minimal energy, which have to respect the topological restrictions and the boundary conditions. It will be interesting to investigate whether the prediction of these magnetic currents is in contradiction to experimental observations. We will see that these magnetic currents do not lead to additional forces acting on electric currents besides the well known Coulomb and Lorentz forces from electric and magnetic field strengths.

%------------------------------------------------------------------------------
\subsection{Coulomb and Lorentz forces}\label{sec:Forces}
%------------------------------------------------------------------------------
Due to the joint description of charges and electromagnetic fields by $Q(x)$, the interaction of solitons is a consequence of topology~\cite{Chan:1993se}. The forces between charges and electromagnetic fields are internal forces and the total force density is vanishing, see Eq.~(\ref{Kraftdichte}). The electrodynamic limit provides the possibility to separate charges and their fields.

In the electrodynamic limit the energy-momentum tensor~(\ref{DefEMT}) tends to
\begin{equation}\label{ExpEMomTen}
\Theta^\mu_{\;\nu}(x)\to T^\mu_{\;\nu}(x)\ist{DefEMT}
-\frac{1}{\mu_0}{\hspace{0.5mm}^\star}\hspace{-0.5mm}f_{\nu\sigma}(x)
 {\hspace{0.5mm}^\star}\hspace{-0.5mm}f^{\mu\sigma}(x)-\mathcal L_{\rm ED}(x)\,\delta^\mu_\nu,
\end{equation}
with components of the same form as the symmetrised energy-momentum tensor in Maxwell's theory
\begin{eqnarray}
T^0_{\;0}&=&\mathcal H\ist{ExpEMomTen}\frac{\varepsilon_0}{2}
 \left[\mathbf E^2+c_0^2\mathbf B^2\right],\\
T^0_{\;i}&\ist{ExpEMomTen}&-c\varepsilon_0\,(\mathbf E\times\mathbf B)_i,\\
T^i_{\;j}&\ist{ExpEMomTen}&\varepsilon_0\left[E_iE_j+c_0^2B_iB_j-\frac{\delta^i_j}{2}
  (\mathbf E^2+c_0^2\mathbf B^2)\right].
\end{eqnarray}
We can use Newton's third axiom where the reaction to the force density of electromagnetic fields is the force density $f^\mu_\mathrm{e}$ on charges. Then charges appear as external sources
\begin{equation}\begin{aligned}\label{forcedensityderivation}
f^\mu_\mathrm{e}&:\ist{Kraftdichte}-\partial^\nu T^\mu_{\;\nu}\ist{ExpEMomTen}
\frac{1}{\mu_0}\partial^\nu\left({\hspace{0.5mm}^\star}\hspace{-0.5mm}
f_{\nu\rho}{\hspace{0.5mm}^\star}\hspace{-0.5mm}f^{\mu\rho}\right)
-\frac{1}{4\mu_0}\partial^\mu({\hspace{0.5mm}^\star}\hspace{-0.5mm} f_{\rho\nu}
{\hspace{0.5mm}^\star}\hspace{-0.5mm}f^{\rho\nu})=\\
&=\frac{1}{\mu_0}
[\underbrace{\partial^\nu{\hspace{0.5mm}^\star}\hspace{-0.5mm}f_{\nu\rho }}
_{\frac{1}{c_0} g_\rho}{\hspace{0.5mm}^\star}\hspace{-0.5mm}f^{\mu\rho}
+\underbrace{{\hspace{0.5mm}^\star}\hspace{-0.5mm}f_{\nu\rho}
\;\partial^\nu{\hspace{0.5mm}^\star}\hspace{-0.5mm}f^{\mu\rho}}_{
-{\hspace{0.5mm}^\star}\hspace{-0.5mm}f_{\nu\rho}
\partial^\rho{\hspace{0.5mm}^\star}\hspace{-0.5mm}f^{\mu\nu}}
+\frac{1}{2}{\hspace{0.5mm}^\star}\hspace{-0.5mm}f_{\nu\rho}
\;\partial^\mu{\hspace{0.5mm}^\star}\hspace{-0.5mm}f^{\rho\nu}]=\\
&\ist{magcur}\frac{1}{\mu_0 c_0}\underbrace{
{\hspace{0.5mm}^\star}\hspace{-0.5mm}f^{\mu\rho} g_\rho}_{0}
+\frac{1}{2\mu_0}{\hspace{0.5mm}^\star}\hspace{-0.5mm}f_{\nu\rho}
[\underbrace{\partial^\nu{\hspace{0.5mm}^\star}\hspace{-0.5mm}f^{\mu\rho}
+\partial^\rho{\hspace{0.5mm}^\star}\hspace{-0.5mm}f^{\nu\mu}
+\partial^\mu{\hspace{0.5mm}^\star}\hspace{-0.5mm}f^{\rho\nu}}_{
-\mu_0\epsilon^{\mu\nu\rho\sigma} j_\sigma}]=\\
&\iist{inhmax}{obsEOM}-\frac{1}{2}\epsilon^{\mu\nu\rho\sigma}
{\hspace{0.5mm}^\star}\hspace{-0.5mm}f_{\nu\rho} j_\sigma 
=f^{\mu\sigma} j_\sigma.
\end{aligned}\end{equation}
The components of the force density on charges read
\begin{eqnarray}\label{forcedensity0}
&&f^0_\mathrm{e}=\frac{1}{c_0}\mathbf j\mathbf E,\\\label{forcedensity3}
&&\mathbf{f}_\mathrm{e}=\rho\mathbf E+\mathbf j\times\mathbf B.
\end{eqnarray}
Eq.~(\ref{forcedensity0}) describes the loss of power density of the field and the corresponding increase for charges. The spatial force density~(\ref{forcedensity3}) includes Coulomb- and Lorentz forces acting on point-like electric charges.

The above relations show that the electrodynamic limit of the soliton model corresponds to Maxwell's electrodynamics (MEdyn), with the distinction that only integer multiples of the electric charge unit are allowed. This restriction leads to the appearance of magnetic currents which do not directly contribute to the forces on electric currents and thus are only internal currents. They are not quantised and influence electric charges only via their electric and magnetic fields.

From MEdyn we know that conserved electric currents are related to a U(1)-gauge invariance.

%------------------------------------------------------------------------------
\subsection{U(1) gauge invariance}\label{Sec-U1}
%------------------------------------------------------------------------------
According to Eq.~(\ref{CurvWu}), $\vec R_{\mu\nu}$ is a vector in the su(2) algebra, parallel to $\vec n$. \label{U1gaugeInv}This gives rise to the emergence of a local U(1) gauge symmetry. A gauge transformation corresponds to a x-dependent rotation of the three basis vectors $\sigma_i$ of the su(2) algebra under rotations $\vec\omega(x)=\omega(x)\vec n$ around $\vec n$ by an arbitrary angle $\omega(x)$
\begin{equation}\label{sigmaRot}
\vec\sigma\;\to\;\vec\sigma(\vec\omega(x))
:=\mathrm e^{\omega(x)\vec n\times}\vec\sigma.
\end{equation}
For infinitesimal coordinate shifts these transformations read
\begin{equation}\label{sigmaRotdiff}
\vec\sigma(\vec\omega(x+\mathrm dx))\ist{sigmaRot}
\mathrm e^{\mathrm dx^\mu\partial_\mu\omega\,\vec n\times}\vec\sigma(\vec\omega(x)).
\end{equation}
They have to be compensated by an additional rotation of $\vec n$ vectors. The affine connection~(\ref{derivativeQ}) is the rotational angle transforming $\vec n(x)$ to $\vec n(x+\mathrm dx)$
\begin{equation}\begin{aligned}\label{nrotieren}
&\vec n(x+\mathrm dx)\ist{derivativeQ}
\mathrm e^{\mathrm dx^\mu\vec\Gamma_\mu\times}\vec n(x)
=\vec n(x)+\mathrm dx^\mu\vec\Gamma_\mu\times\vec n,\\
&\vec n(x+\mathrm dx)=\vec n(x)+\mathrm dx^\mu\partial_\mu\vec n.
\end{aligned}\end{equation}
Comparing the two equations, we see that $\vec\Gamma_\mu$ has to respect
\begin{equation}\label{BedAnGamma}
\vec\Gamma_\mu\times\vec n\ist{nrotieren}\partial_\mu\vec n.
\end{equation}
Eq.~(\ref{ConWu}) satisfies this condition. But $\vec\Gamma_\mu$ is fixed by Eq.~(\ref{BedAnGamma}) up to a component parallel to $\vec n$ only. This freedom offers the possibility to compensate the basis transformation~(\ref{sigmaRotdiff}) by the shift
\begin{equation}\label{GammaTrafo}
\vec\Gamma_\mu\ist{ConWu}\vec n\times\partial_\mu\vec n\;\to\;
\vec\Gamma_\mu^\prime:=\vec n\times\partial_\mu\vec n+\partial_\mu\omega\,\vec n.
\end{equation}
With this transformation, algebra valued fields get invariant under basis transformations.

As Eq.~(\ref{GammaTrafo}) reminds us the affine connection is not a tensor.

According to the above derivation the $\vec n$-field and therefore the coordinates~(\ref{CurvWu}) of the curvature tensor $\vec R_{\mu\nu}$ are invariant under local gauge transformations, realised in this model as basis transformations in the tangential space of $\mathbb S^2$. This $\mathbb S^2$ is the parameter space of the soliton field $Q(x)$ in the electrodynamic limit~(\ref{Qlimit}).

A further important application of the electrodynamic limit is the sector where only electromagnetic waves are present.

%------------------------------------------------------------------------------
\subsection{Goldstone bosons}\label{Sec-Gold}
%------------------------------------------------------------------------------
The minimum $q_0=0$ of the potential in the Lagrangian~(\ref{Lagr4D}) is degenerate at a two-dimensional manifold. Thus, the vacuum has broken symmetry and the model has two types of Goldstone bosons. The Lagrangian reduces to the form~(\ref{EDLagrangian}). Up to the different sign, this Lagrangian is formally identical to the Lagrangian of Maxwell's electrodynamics with 2*4 physical dofs of the wave-field in radiation gauge when the time component and the longitudinal component of the gauge field are removed. In distinction, in the present model the $\vec n$-field has two dofs only. We describe them by two spherical angles  $\theta(x)$ and $\phi(x)$. In the electrodynamic limit and in the absence of singularities of the $\vec n$-field there are no charged particles present and we can assume that the field at spatial infinity ``$\infty$'' is independent of the direction. The vacuum has broken symmetry. As a consequence of the  Hobart-Derrick theorem~\cite{Hobart:1963rh,Derrick:1964gh}, pure regular $\vec n$-field configurations moving with velocities smaller than the velocity of light $c_0$ are unstable. For velocities $c_0$ they have constant action in time and one can try to describe electromagnetic waves~\cite{borisyuk:2007bd}. Due to the isomorphism $\mathbb R^3 \cup \infty \sim \mathcal{S}^3$ and the topological relation $\pi_3(\mathcal{S}^2)=\mathbb{Z}$ there is an additional quantum number for the $\vec n$-field, the Hopf number or Gauß linking number $v$ of fibres $\mathcal F$ defined by $\vec n_\mathcal{F}=$const, thus by certain values $\theta_\mathcal{F}$ and $\phi_\mathcal{F}$. Fibres are in general closed lines and for different fibres $\mathcal C_1$ and $\mathcal C_2$ we can determine a linking number $v$ by the famous double integral of Carl Friedrich Gauß
\begin{eqnarray}\label{GaussDoppelInt}
v\,=\,\frac{1}{4\pi}\oint_{\mathcal C_1}\oint_{\mathcal C_2}
\frac{\mathbf r_1-\mathbf r_2}{|\mathbf r_1-\mathbf r_2|^3}
\cdot(\mathrm d\mathbf r_1\times\mathrm d\mathbf r_2),
\end{eqnarray}
With $v=\pm1$ we identify right and left polarised photons. $v$ can be modified by the interaction of photons with solitons. The physical equivalent of $v$ is the number of photons $n_\gamma$ in a field configuration.

As described in Refs.~\cite{Jech2014,Faber:2017uvr} $v$ can be computed also by one-dimensional integrals from the behaviour of neighbouring fibres. In such a determination enters the geometry of $S^2$ in an interesting way.

% xxx details of one dimensional integrals

Crossing solitons in arbitrary direction rotates the spatial Dreibein by $2\pi$ with left or right chirality as depicted in Tab.~\ref{TabSign}. Scattering, emission and absorption processes of photons by solitons are able to modify the chiralities of both types of particles. These geometrical processes need detailed investigations.

Much work on $\mathbb S^2$-fields with different Lagrangians has already be done by Ferreira et al., see e.g.~\cite{Ferreira:2009} and references therein. A recent article about the closely related field of Topological photonics is~\cite{Shen:2021}.

%------------------------------------------------------------------------------
\section{Open questions and Conjectures}\label{Sec-Ideen}
%------------------------------------------------------------------------------
In this section we discuss some features of the model which need future detailed investigations or where in the moment we can rely on speculations only.

%------------------------------------------------------------------------------
\subsection{Running of the coupling}\label{Sec-Running}
%------------------------------------------------------------------------------
The usual conclusion from high energy scattering experiments is that electrons are point-like. This could counteract this model, since this model suggests a finite size of static electrons of the order of the classical electron radius $r_\mathrm{cl}$.

In order to learn from experiments about the property of electrons at distances of $r_\mathrm{cl}$ and smaller we need high energies. A characteristic relation between radius and energy is $\hbar c_0\approx200~\textrm{MeV fm}$. 200 MeV electrons are sensible at distances of fm, the 100 GeV electrons of LEP up to distances of 2 attometers.

The solitons of this model are relativistically covariant. A moving soliton is Lorentz compressed. At LEP, electrons reached a $\gamma$ factor of $2\cdot10^5$ and therefore could approach another electron in a head-on collision up to distances of $0.5\cdot10^{-5}\,r_\mathrm{cl}=0.014$~am. The analogous behaviour one can study in the analytical two-particle solutions of the Sine-Gordon model. For $\gamma$ to infinity the distance of closest approach of two solitons approaches zero.

The finite size of solitons of this model leads to a modification of Coulomb's law at distances of the order of the soliton radius. This effect is known in quantum field theories as running of the coupling. The result of the perturbative calculations in QED is the Uehling potential~\cite{peskin1995}. It predicts a strong rise of the effective charge of electrons at the order of a few fm. Since nothing in nature seems to be infinitely large of infinitely small one can see this rise as an indication for an effective size of electrons at this order of magnitude.

The model discussed in this article gets a running of the coupling at the classical level. In due course we will publish a paper about this effect. The numerical calculations were prepared in a paper published recently in Few Body systems~\cite{Anmasser:2021ubc}.

%------------------------------------------------------------------------------
\subsection{Orthogonality of $\mathbf E$ and $\mathbf B$}\label{Sec-perp}
%------------------------------------------------------------------------------
We have seen in Eq.~(\ref{EBNull}) that the conservation of topological charge leads to $\vec{\hspace*{0.5mm}\mathbf E}\vec{\hspace*{0.5mm}\mathbf B}=0$. In the electrodynamic limit we have $\vec R_{\mu\nu}\ist{AbelianFS}R_{\mu\nu}\vec n$ and therefore also $\mathbf E\cdot\mathbf B=0$, as is also implied by Eq.~(\ref{obsEOM}) for non-vanishing magnetic currents. This seems to contradict to the experimental situation, where we have no problems to produce parallel electric and magnetic fields. There is possibly a way-out of this problem, see also the discussion in Ref.~\cite{faber:2002nw}. In the electrodynamic limit the basic field is the unit-vector field $\vec n(x)$. A constant $\vec n$-field corresponds to zero field strength. One can get the electric field by an $\vec n$-field rotating in space. Due to the topological restriction, a homogeneous electric field can be produced only in a finite spatial region. The solution, given in Ref.~\cite{faber:2002nw} looks more complicated than in Maxwell's theory, but it agrees better to the experimental situation with capacitor plates of finite size and finite charge and fields whose homogeneity is only approximate. Magnetic fields originate from non-vanishing time-components of the connection and thus from the rotations of $\vec n$-vectors in time. Electric currents in a wire, the source of magnetic fields, correspond to hoppings of electrons from one ion to the next in a very short time, a strong time-dependence of $\vec n$-fields is necessary to produce a reasonable magnetic field.

The orthogonality of $\mathbf E$ and $\mathbf B$ follows from the interpretation of the field strength as curvature of the unit vector field $\vec n$, see Eq.~(\ref{CurvWu}). It is not derived from the Lagrangian and thus not a consequence of the dynamics. It is valid at the level of distances between elementary charges. This leads to the idea, that the orthogonality is present on a microscopic level and parallelity may be achieved on a macroscopic time-averaged level.

%------------------------------------------------------------------------------
\subsection{Quantum effects}\label{Sec-fluct}
%------------------------------------------------------------------------------
The model formulated in Sect.~\ref{Sec-Formulation} is a classical model. Since around 100 years we know that nature in atomic and subatomic physics is dominated by quantum effects. Under the present paradigm of quantum field theory one would include quantum effects by integrations over all possible quantum fluctuations. In Ref.~\cite{Faber:2008hr} it was shown that such integrations lead to diverging integrals. In distinction to Maxwell's electrodynamics (MEdyn) the presented model is from the beginning finite. Therefore, there seems no need of regularisation and renormalisation. If the model is of some applicability, there must be another method to include quantum fluctuations and interference, encoded in the path integral formulation of quantum mechanics and quantum field theory by its complex Boltzmann factor  $\mathrm e^{\mathrm iS}$, where $S$ is the action attributed to field configurations. There is some freedom to achieve this goal since nobody knows how nature does, in order to produce the effects which are perfectly described by quantum mechanics.

There are hundreds of books and thousands of articles about the riddles and the philosophical implications of quantum mechanics, but there are only few ideas suggesting mechanisms. I know of only one experiment which may give an idea for a mechanism. This is Couder's experiment~\cite{Couder2005} of bouncing oil drops which, as I think, supports De Broglie's idea~\cite{Broglie:1927} of pilot waves. The bouncing drops are interacting with waves and produce waves themselves~\cite{Bush2015}. In the critical double slit experiments, the waves pass both slits, whereas every particle passes only one slit. The particle trajectories are influenced by the interfering waves and lead to an interference pattern at the detectors.

Can we find such a mechanism in our model? Quantum mechanical fluctuations are action fluctuations and not energy fluctuations. This could mean, the energy is transferred for a short moment and then again subtracted. Some type of waves, unknown in MEdyn, passing a particle could possibly provide such a phenomenon.

There are two types of fields in our model which are unknown in MEdyn, magnetic currents $g^\mu$ and waves in the angular parameter $\alpha$ of the $Q$-field
\begin{equation}\label{Qnalpha}
Q\ist{unitquaternions}q_0-\mathrm i\vec q\vec\sigma\ist{nalpha}\cos\alpha-\mathrm i\vec\sigma\vec n\sin\alpha.
\end{equation}
$\alpha$ is related to the rotational angle $\omega$ of Dreibeins in space by $\vec\omega=2\vec\alpha$. Magnetic currents are non-vanishing sources of magnetic fields~\cite{faber:2002nw}. As was discussed in Sect.~\ref{sec:Forces} magnetic currents do not contribute to Coulomb and Lorentz forces, see Eq.~(\ref{forcedensity3}) and therefore do not dirctly influence the motion of charged particles. The appearance of magnetic currents is a result of the non-Abelian nature of the soliton field. Due to the topological restrictions electromagnetic waves may be accompanied by magnetic currents, which in the vacuum propagate with the velocity of light in the direction given by the Poynting vector. The magnetic currents coming from the soliton description are different from the Dirac magnetic currents, because they are not quantised and cannot be associated with massive magnetic charges. They are suppressed by the request of minimal energy and seem rather tiny deviations from the solutions of the homogeneous Maxwell equations, in case these can not be fulfilled due to the topological restrictions of the $\vec n$-field.  Since magnetic currents seem not to have measurable consequences, they should be ruled out as guiding waves.

\begin{wrapfigure}[9]{R}{5cm}
\vspace{-8mm}
\centering
\includegraphics[scale=0.4]{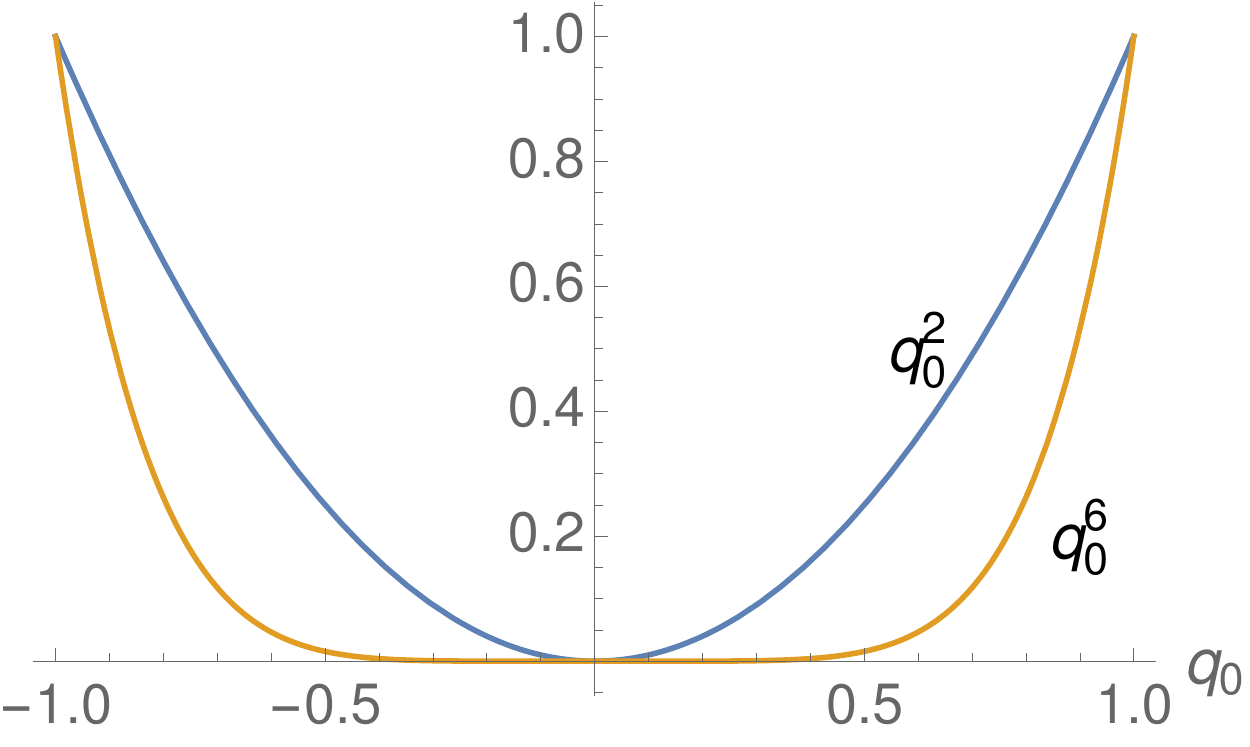}
\caption{Observe the flat region of the function $q_0^6$.}
\label{fig:potential}
\end{wrapfigure}

$\alpha$-waves are governed by the shape of the potential term $\mathcal L_\textrm{pot}$ in the Lagrangian~(\ref{Lagr4D}) with its unusual power $q_0^6$, see Fig.~\ref{fig:potential}. In Ref.~\cite{Faber:2008hr} it was shown that a $q_0^2$ dependence of the potential term would lead to vibrational excitations of solitons. The aim to compare solitons with the lightest charged fermions is ruling out such excitations. Higher even powers in $q_0$ do not show such vibrations but allow for waves in $q_0=\cos\alpha$. For radial waves a Schrödinger-type equation
\begin{equation}\label{DifEq}
\left[-\frac{1}{2}\Delta+V(r)\right]\eta_k(\mathbf{r})
=\omega^2_k\eta_k(\mathbf{r}).
\end{equation}
with the potential
\begin{equation}\label{potential}
V(r)=\frac{1+3\cos2\alpha}{4r^2} +\frac{15r^2}{2r_0^4}\cos^4\!\alpha
\ist{sincos}\frac{2r_0^4+r_0^2r^2+14r^4}{2r_0^2r^2(r_0^2+r^2)^2}
\end{equation}
was derived which is repulsive everywhere and has no bound states. The spectrum of spherical waves  
\begin{equation}\label{spherical}
\eta_k=\frac{e^{\pm ikr}}{r}\;\qquad \text{with}\qquad
\omega_k^2=\textstyle{\frac12}k^2.
\end{equation}
is continuous.

In scattering processes between photons lumps of non-vanishing $q_0(x)\ist{nalpha}\cos\alpha$ are produced. As mentioned in Sect.~\ref{Sec-Gold} photons are characterised by a field of purely imaginary unit quaternions, see Eq.~(\ref{Qlimit}). In scattering processes two such fields $Q_1(x)$ and $Q_2(x)$ have to be multiplied
\begin{equation}\label{photcross}
Q_1(x)Q_2(x)=[-\mathrm i\vec\sigma\vec n_1(x)][-\mathrm i\vec\sigma\vec n_2(x)]
=-(\vec n_1\vec n_2)-\mathrm i\vec\sigma(\vec n_1\times\vec n_2)
\end{equation}
leading to a non-vanishing real part $\cos\alpha=-\vec n_1\vec n_2$ and therefore to a small contribution to the potential energy, to a tiny massive lump acting like a small particle.

An oscillation of non-vanishing $q_0$ in some direction $\vec n_o$ could bounce way and back in the flat region of the $q_0^6$-potential. If such an oscillation approaches a soliton, due to its tiny mass contributions it should shift the soliton center depending on the angle between $\vec n_o$ and the solitons $\vec n$-vector in this spatial region. Due to the above mentioned oscillation in the $q_0^6$-potential the soliton center should move alternating in forward and backward direction, i.e. adding an subtracting energy and momentum in short sequence. This would maintain energy and momentum constant, but increase the action. It will be interesting to investigate in more detail whether such $\alpha$-waves could provide a model for the physical nature of a guiding field. In diffraction experiments the diffracted field would exist independent of the presence of solitons and the solitons could maintain their corpuscular nature all along the experiment as the authors of Ref.~\cite{Pena:2015mf} expect for the zero-point fluctuations as guiding field.

$\alpha$-waves could be a candidate for dark matter. Most of the community is looking for dark particles. Up to now nobody found dark particles. I think, dark matter escapes our detectors since it does not have the form of particles and it does not modify the kinetic energy of particles. $\alpha$-waves do not have particle structure. Due to their contribution to the potential they are rather tiny lumps of not quantised mass.

%------------------------------------------------------------------------------
\subsection{A possible mechanism of cosmic inflation}\label{Sec-infl}
%------------------------------------------------------------------------------
Some cosmological models assume an inflationary epoch~\cite{Guth:1981} in the early universe, a very short period of exponential expansion of the cosmos, explaining the large-scale structures of our universe, its flatness and the relative homogeneity and isotropy of the cosmic microwave background.  Our model provides a possible mechanism for inflation. A transition from an initial field configuration in the early universe from an initial stationary state with positive energy density $\mathcal H^\textrm{before}$
\begin{equation}\label{before}
Q(x)\equiv1\;\Rightarrow\;\Lambda\ist{Lagr4D}\frac{1}{r_0^4}\;\Rightarrow\;
\mathcal H^\textrm{before}:=-\mathcal  L_\textrm{pot}^\textrm{before}\ist{Lagr4D}
\frac{\alpha_f\hbar c_0}{4\pi r_0^4}
\end{equation}
to a stable vacuum with vanishing energy density $\mathcal H^\textrm{after}$
\begin{equation}\label{after}
\hspace{-7mm}Q(x)\equiv-\mathrm i\sigma_3\;\Rightarrow\;\Lambda\ist{Lagr4D}0\;\Rightarrow\;
\mathcal H^\textrm{after}:=-\mathcal  L_\textrm{pot}^\textrm{after}\ist{Lagr4D}0
\end{equation}
after inflation.~\footnote{Despite the vanishing energy-momentum tensor in the absence of solitons, there is a running of the coupling due to the modification of Coulomb's law at distances of the order of the soliton radius, as will be discussed in detail in a forthcoming article. This is in contradistinction to quantum electrodynamics, where quantum corrections lead to a running coupling described by a non-vanishing beta-function. There due to the proportionality between beta-function and the value of the trace of the energy-momentum tensor, see Eq.(19.157) of Peskin-Schröder~\cite{peskin1995}, a vanishing trace would prevent the running of the coupling.} This transition would have released the energy density 
\begin{equation}\label{Efrei}
\mathcal H^\textrm{before}-\mathcal H^\textrm{after}\iist{before}{after}
=\frac{\alpha_f\hbar c_0}{4\pi r_0^4}\ist{r0m2}4.8~\textrm{keV/fm}^3=7.7\cdot 10^{29}J/\textrm m^3.
\end{equation}

%------------------------------------------------------------------------------
\subsection{Cosmological constant}\label{Sec-Lambda}
%------------------------------------------------------------------------------
1917 Einstein has introduced  a cosmological constant in general relativity to compensate the attractive interaction of the mass density and to get stationary solutions of the field equations. In cosmological models, like the $\Lambda$CDM model~\cite{turner:2021}, the cosmological constant is related to an energy density of the vacuum explaining the observed cosmic expansion. The Planck collaboration gives in its 2018 results~\cite{Planck:2018} a cosmological energy density of $0.69\,\rho_\mathrm{crit}$, where the critical energy density is $4.9\,\frac{\textrm{GeV}}{\textrm m^3}$.

The potential term $\mathcal L_\mathrm{pot}$ in the Lagrangian~(\ref{Lagr4D}) acts like a cosmological function, contributing to the energy density of the universe. Due to the application of the Hobart-Derrick theorem a quarter of the restenergy of solitons is contributed from the potential energy $H_\mathrm{pot}$, see Eq.~(\ref{EnergieBez}). If this ratio would also be valid for nucleons an average density of nucleons of 14.5 nucleons per m$^3$ would agree with the measured energy density. In comparison, the prediction of field theory deviates many orders of magnitude - a still unsolved problem.

As discussed in Sect.~\ref{Sec-fluct}, due to their contributions to the potential energy $\alpha$-waves would also participate to the cosmological constant, the average of the cosmological function.

%------------------------------------------------------------------------------
\section{Conclusions}\label{Sec-Zusammenfassung}
%------------------------------------------------------------------------------
\subsection{Comparison to other models}\label{Sec-Vergleich}
%------------------------------------------------------------------------------
The model discussed in this article has many features in common with other well-known models. The Lagrangian defined in Eq.~(\ref{Lagr4D}) can be seen as a generalisation of the Sine-Gordon model~\cite{remoissenet:2003}, a model in 1+1D with one field degree of freedom (dof), generalised to a field in 3+1D with three dofs.

Up to the difference between the groups SU(2) and SO(3) the dofs are the same as those of the Skyrme model~\cite{Skyrme:1958vn,Skyrme:1961vq}. The difference is in the terms compressing solitons, in the Skyrme model the compressing term has two derivatives, in the Lagrangian~(\ref{Lagr4D}) it is substituted by the term $\mathcal  L_\textrm{pot}$ without derivatives.

Dirac~\cite{dirac:1931kp,dirac:1948um} suggested to introduce magnetic monopoles in Maxwell's electrodynamics (MEdyn) by defining a line-like singularity, the well-known Dirac string. Wu and Yang~\cite{wu:1975vq,wu:1976qk} succeeded in removing this singular line using a normalised three-component scalar field $\vec n$ to describe the field of monopoles, but they did not remove the singularity in the center of the monopoles.

'T Hooft~\cite{tHooft:1974kcl} and Polyakov~\cite{Polyakov:1974ek} investigated monopoles in gauge field theories and identified monopoles in the Georgi-Glashow model. This model has a triplet of gauge fields with $3\cdot4$ dofs and a triplet of scalar (Higgs) fields, in summary 15 dofs. The energy of these monopoles were estimated with the order of the mass of the W-boson multiplied with the inverse of Sommerfeld's fine structure constant. As was shown in~\cite{Polyakov:1976fu},\cite{DIETZ:1980536} and \cite{Manton:1977er} in the so-called BPS-limit~\cite{Bogomolnyi:1976kr},\cite{Prasad:1975kr} of the massless Higgs boson, monopoles and antimonopoles keep interacting with each other Coulomb-like. However, the force between two monopoles or two antimonopoles vanishes, whereas the interaction-strength between a monopole and an antimonopole doubles in comparison with the limit where the Higgs boson is much heavier than the W-bosons of the 3D Georgi-Glashow model.

An attempt to construct the Standard Model with Monopoles was undertaken by Vachaspati~\cite{Vachaspati:1995yp}. However, in that article, which is aimed at discussing the algebraic properties of possible mappings between particles and monopoles, dynamical features were set aside.

We describe particles and their fields with the same dofs, in analogy to the Sine-Gordon and the Skyrme model. Only three dofs of an SO(3)-field are necessary to describe charges and their fields. Since these three dofs can be interpreted as orientations of spatial Dreibeins we can ascribe these properties to space-time. In this way we would succeed to describe the two long-range interactions, electrodynamics and gravitation, by the properties of space-time only.

An important test of the applicability of this model is an investigation of the differences to MEdyn.

%------------------------------------------------------------------------------
\subsection{Comparison to Maxwell's electrodynamics}\label{Sec-EMVergleich}
%------------------------------------------------------------------------------
The model describes charges with long range forces as we have in Maxwell's electrodynamics (MEdyn). As already mentioned, this model can not substitute Maxwell's electrodynamics and the SM. The purpose of this chapter is to try to enumerate where this model differs from these theories, where it may be successful and where it may fail.

Particles are characterised by topological quantum numbers, only artificially they can be separated from their fields. Such a separation leads to the well-known singularities of point-like charges.

The following properties agree with MEdyn:
\begin{enumerate}
\item The Lagrangian~(\ref{Lagr4D}) is Lorentz covariant, thus the laws of special relativity are respected.
\item Charges have Coulombic fields fulfilling Gaußes law~(\ref{inhmax}).
\item Charges interact via $\frac{1}{r^2}$ electric fields~(\ref{EFieldStrengthHedgeHog}), they feel Coulomb and Lorentz forces, see Sect.~\ref{sec:Forces}.
\item A local U(1) gauge invariance is respected, see Sect.~\ref{Sec-U1}.
\item There are two dofs of massless excitations for photons, see Sect.~\ref{Sec-Gold}.
\end{enumerate}

The critical questions concern the differences to MEdyn. Several differences agree with the experiment, others have to be investigated in detail, whether they differ only superficially or concern deep discrepancies. In distinction to MEdyn we find the following properties:
\begin{enumerate}
\item Electric charges are quantised, see Sect.~\ref{Sec-Ladung}, like the magnetic charges of Dirac monopoles. Charge is a topological quantum number.
\item The topological construction explains the astonishing mirror properties of particles and antiparticles~\cite{Borchert2022}, see Sect.~\ref{Sec-Qnum}.
\item The mass of solitons is completely due to field energy, see Eq.~(\ref{totEneMonop}), and finite.
\item The self-energy of charges is finite and does not need regularisation and renormalisation, see Eq.~(\ref{EneMonopFormel}).
\item Charges and their fields are described by the same dofs, see Sect.~\ref{Sec-Formulation}, by SO(3) rotational degrees of freedom. The calculations are simplified using the dofs, see Eq.~(\ref{unitquaternions}), of the double covering group SU(2) of SO(3).
\item These SO(3) dofs can be interpreted as orientations of spatial Dreibeins, see Sect.~\ref{Sec-Formulation}. Thus, besides the properties of space-time no additional fields are necessary to describe electro-magnetism and gravitation.
\item Gauge symmetry is explained as a geometrical phenomenon, as basis changes in the tangential manifold on $S^3$, see the paragraph between Eqs.~(\ref{RAllg}) and (\ref{FDvonR}).
\item Spin appears with usual quantisation properties known from quantum mechanics, see Sect.~\ref{Sec-Spin}, and respects the combination rules of representations of SU(2). Spin is attributed to field configurations with values in the group manifold, whereas in quantum field theories it is usually attributed to algebra valued fields acting on vectors in the corresponding representation.
\item There are four basic configurations of solitons, see Tab.~\ref{TabSign}, with the quantum numbers~(\ref{TeilchenSpin}) of components of Dirac spinors.
\item Solitons and antisolitons have opposite internal parity, see Tab.~\ref{TabSign}, as is well-known for the description of electrons and positrons with the Dirac equation.
\item Solitons are characterised by a chirality quantum number which can be related to the sign of the magnetic quantum number.
\item Spin contributes to angular momentum due to internal rotations of soliton centers, see Fig.~\ref{rotating}.
\item The canonical energy-momentum tensor~(\ref{DefEMT}) does not need additional symmetrisation. It is automatically symmetric.
\item Due to the dual representation~(\ref{FDvonR}) of the field strength tensor static charges are described by the spatial components of vector fields~(\ref{sphConCof}). For time-dependent phenomena, like moving charges, electric currents and magnetic fields, non-vanishing time-components of the connection $\vec\Gamma_\mu$ are necessary, see Eqs.~(\ref{RSU2},\ref{FDvonR}).
\item The finite size of solitons leads to a modification of the Coulomb's law at distances of the order of the soliton radius~\cite{Anmasser:2021ubc}. This effect is known in quantum field theories as running of the coupling, see Sect.~\ref{Sec-Running}.
\item The local U(1) gauge invariance is explained by the free choice of the bases in the tangential spaces of the $\mathbb S^2$ manifold, see Sect.~\ref{Sec-U1}.
\item Photons are characterised by a topological quantum number, the Gauß\-ian linking number~(\ref{GaussDoppelInt}) of fibres of the $\mathbb S^2$-field, the soliton field in the electrodynamic limit.
\item The photon number can be modified by interaction of photons with charges, see Sect.~\ref{Sec-Gold}.
\end{enumerate}

The following properties differing from MEdyn need deeper investigation and may possibly differ from experiments with electrons and photons.
\begin{enumerate}
\item Spin and magnetic moment are dynamical properties. They are consequences of movements of solitons or time-dependent field variations in their surroundings.
\item Electric and magnetic field vectors are perpendicular to each other, see Sect.~\ref{Sec-perp}.
This is in obvious contradiction to experimental realisations of parallel electric and magnetic fields. An excuse may be, that this property may be realised on a microscopic level and allows for parallelity of these fields on a macroscopic level.
\item The existence of unquantised magnetic currents is allowed. These currents do not directly appear in Lorentz forces. They contribute only via their magnetic fields.
\item $\alpha$-waves with oscillating values of $q_0=\cos\alpha$ feel non-vanishing values of the potential energy density and contribute therefore to matter density. Such non-topological excitations were not directly detected. If they exist, they contribute to dark matter, see Sect.~\ref{Sec-fluct}.
\item $\alpha$-waves lead to additional forces on particles and are a possible origin of quantum fluctuations.
\item The potential term allows for a mechanism of cosmic inflation.
\item The potential term contributes to dark energy.
% xxx\item Aharonov-Bohm like water vortex
\end{enumerate}
This model tries to give an idea about a possible direction of future investigations. One should not assume that it can provide a final answer to all questions which are left open by quantum theory and quantum field theory. Both theories have a tremendous success. It is obvious that the model cannot compete with their excellent and precise predictions and the century-long efforts of numerous scientists to describe the properties of nature in the most fundamental domain. But this does not mean that one should not think about scenarios which finally could lead to a deeper understanding of the mechanisms in nature. This could allow us to answer some questions, left open by quantum mechanics and quantum field theory. It could help to answer how nature works to produce the interesting riddles which we could not solve yet.

Some of the questions posed by this model concerning electrodynamics, quantum theory and quantum field theory are enumerated in the section with critical remarks. Many may still be missing. Essential open problems are also those related to the three other fundamental forces, gravitational, strong and weak interactions.

If an extension of this model is of some relevance for describing the properties of nature, one could summarise its philosophy in: Particles are properties of regions of space, characterised by topological quantum numbers. Therefore I would like to call it: ``Model of topological particles''.

%-------------------------------------------------------------------------------
\section{Aftermath}
%-------------------------------------------------------------------------------
In its fundaments, physics is measurements of distances of objects and times of events, and explanation of their relations.\\ This may indicate, that\\
\hspace*{10mm}Physics is geometry and not algebra.\\
\hspace*{10mm}Finally, one should use the algebra to describe the geometry.\\[5mm]
\noindent General Relativity:\\
\hspace*{5mm}Wheeler: ``Spacetime tells matter how to move;\\
\hspace*{25mm}matter tells spacetime how to curve.''\cite{Wheeler1998}\\[1mm]
My addition for Electrodynamics:\\
$\cdots$ Charges and electromagnetic fields tell space how to rotate.\\

%-------------------------------------------------------------------------------
\section*{Acknowledgement}
%-------------------------------------------------------------------------------
I want to thank Jarek Duda for his interest in the model and his support to make it more public in the community. To the anonymous referees of this article I am grateful for their constructive comments.

\bibliographystyle{utphys}
\bibliography{geometricmodel}

\end{document}